# VLM-VPI: A Vision–Language Reasoning Framework for Improving Automated Vehicle–Pedestrian Interactions

Qingwen Pu[a], Kun Xie[a,*], and Yuxiang Liu[a]

[a] *Transportation Informatics Lab, Department of Civil and Environmental Engineering, Old Dominion University, Norfolk, VA 23529, United States*
[*] *Corresponding Author; Email: kxie@odu.edu*

## Abstract

Autonomous driving systems often infer pedestrian yielding behavior from geometric and kinematic cues alone, limiting their ability to reason about visual scene context and age-dependent behavioral variability. This limitation can produce delayed interventions in safety-critical encounters and unnecessary braking in benign interactions. This work introduces Vision–Language Model-based Vehicle–Pedestrian Interaction (VLM-VPI), a multimodal reasoning framework for pedestrian intent understanding and yielding-aware control in autonomous driving. The system combines three components: a multimodal perception layer that captures visual and kinematic observations, a reasoning layer that uses Qwen3-VL 8B for visual scene understanding and GPT-OSS 20B for few-shot intent reasoning, and a tiered safety controller that applies age-specific braking margins for children, adults, and seniors. In 112 CARLA scenarios, VLM-VPI achieves 92.3% intent classification accuracy, outperforming a rule-based baseline (78.4%), supervised trajectory models (73.5-82.4%), and a zero-shot LLM configuration (88.4%). Validation on 24 real-world PIE scenarios yields 87.5% accuracy, indicating functional sim-to-real transferability. Across 200 simulation cases, VLM-VPI reduces the false-alarm rate from 7.4% to 2.8% and mean intersection traversal time from 13.5 s to 11.8 s. Conflict occurrences decrease from 124 to 33, while mean minimum time-to-collision improves from 1.92 s to 4.47 s. Demographic-adaptive control further reduces conflicts by 60% for children and 54.5% for seniors compared with uniform control. These results show that an explicit vision-language reasoning layer can improve both safety and efficiency by linking pedestrian intent, demographic context, and vehicle control decisions.

## 1. Introduction

Despite remarkable advances in perception, localization, and motion planning, automated vehicles (AVs) continue to face fundamental challenges in safety-critical interactions with pedestrians at urban intersections, where collision risk arises not from sensor failure but from systematic misinterpretation of human intent (Chougule *et al.* 2024). Industry data reveal that pedestrian-involved incidents in AV testing disproportionately occur in scenarios where pedestrians were detected but their crossing intentions were misjudged (Hulse 2023), leading to either delayed emergency braking or unnecessary interventions that disrupt traffic flow (Almeida *et al.* 2026). These failures are particularly concerning for vulnerable populations. Data from the National Highway Traffic Safety Administration (Center 2025) show that pedestrians aged 65 and older account for a disproportionately high share of pedestrian fatalities, while 17% of traffic fatalities among children aged 14 and younger involve pedestrians. These statistics indicate that failures in intent understanding do not merely reduce system efficiency, but directly translate into elevated fatality risks (Godbole *et al.* 2025). This persistent failure exposes a critical limitation: current autonomous driving architectures are effective at detecting and localizing pedestrians in geometric space, yet struggle to infer whether pedestrians will yield—a judgment that depends on latent intent and social reasoning and is not solely determined by observable kinematics (Crosato *et al.* 2024, Mirzabagheri *et al.* 2025).

These failures reflect a mismatch between geometry-driven autonomy architectures and the semantic nature of human pedestrian decision-making (Tzigieras 2025). Current systems operate entirely in geometric and kinematic space, assuming pedestrian behavior is continuous, deterministic, and physics-governed (Prédhumeau *et al.* 2021). However, human crossing decisions occur in semantic and social space, driven by discrete intent states that reverse instantaneously based on belief updates about vehicle approach, right-of-way norms, and perceived danger (Pu *et al.* 2026b). A pedestrian standing motionless may suddenly cross without kinematic warning; conversely, one mid-crossing may abort abruptly upon noticing the vehicle (Wu *et al.* 2023). These categorical transitions are invisible to systems reasoning purely about positions and velocities (Ye *et al.* 2026). Machine learning approaches attempting to learn intent patterns from data confront the same barrier: they learn correlations between observable features and outcomes, not the underlying reasoning process (Barbierato and Gatti 2024). End-to-end policies trained on millions of scenarios still fail catastrophically when deployed in new cultural contexts where yielding norms differ, because they have never learned what yielding means as a social concept (Lake *et al.* 2017). Moreover, both geometric and learning-based frameworks rely on fixed safety thresholds, such as identical braking distances or time-to-collision criteria, for all pedestrians, ignoring that collision risk varies systematically by age (Sheikh and Peng 2025). For example, children exhibit elevated behavioral unpredictability (Xu *et al.* 2023), seniors demonstrate mobility decline with slower crossing speeds (Duim *et al.* 2017), while adults display comparatively more stable crossing patterns (Pitcairn and Edlmann 2000). The fundamental limitation is architectural: current autonomous driving systems lack the capability to recognize demographic categories, reason about population-specific behavioral patterns, and adapt control strategies accordingly (Haghzare *et al.* 2023).

This gap demands a paradigm shift: autonomous vehicles require a semantic reasoning layer that operates between perception and control, translating low-level sensory observations into high-level behavioral understanding of human intent in social space rather than geometric prediction of motion trajectories. Such a layer must interpret pedestrian behavior in terms of social intent rather than kinematic variables and recognize demographic-specific behavioral patterns and collision vulnerabilities to enable age-adaptive safety responses. It must further integrate heterogeneous multimodal data to resolve ambiguity and generate interpretable decisions that can be systematically verified for regulatory certification.

This work introduces Vision–Language Model-based Vehicle–Pedestrian Interaction (VLM-VPI), a multimodal reasoning framework for pedestrian intent understanding and yielding-aware control in autonomous driving. The system integrates three key components: a multimodal perception layer capturing visual and kinematic observations, a reasoning layer employing a vision model (Qwen3-VL 8B) (Bai *et al.* 2025) and a reasoning model (GPT-OSS 20B) (Agarwal *et al.* 2025) to infer pedestrian yielding intent (yielding or non-yielding) and demographic categories (child, adult, or senior) from multimodal inputs, and tiered safety control strategies executing demographic-adaptive braking calibrated to age-specific collision risk profiles. By incorporating real-world behavioral exemplars and demographic-specific safety factors, the framework enables reliable intent inference across diverse pedestrian populations. The core code and several real-world few-shot exemplars are available at: https://github.com/Qpu523/VLM-VPI. The main contributions include:

- Development of a closed-loop autonomous vehicle decision-making and control framework that integrates vision-language reasoning for pedestrian intent understanding, shifting AV decision processes from purely geometric prediction to semantic reasoning in social and behavioral space. This framework bridges the gap between perception and vehicle-level control, enabling interpretable and safety-critical operation in pedestrian interactions.

- Introduction of a few-shot learning methodology that enables LLM reasoning to leverage real-world behavioral patterns through minimal annotated exemplars. This establishes that domain-specific behavioral priors are essential for deploying safety-critical AI systems across different cultural and geographic contexts.

- Implementation of a demographic-adaptive control framework that provides age-aware and differentiated safety margins for children, adults, and seniors, enabling risk-sensitive and context-aware collision avoidance in pedestrian interactions.

## 2. Literature Review

### 2.1. Motion-Based Pedestrian Intention Models

Research on pedestrian safety in automated driving has primarily focused on estimating collision risk through observable kinematics, encompassing surrogate safety metrics, trajectory prediction, and intent classifiers (Ye *et al.* 2026). These approaches are formulated in geometric and motion spaces, attempting to infer pedestrian behavior from spatiotemporal patterns (Dang *et al.* 2025).

Surrogate safety metrics quantify risk through spatial-temporal relationships. Time-to-collision measures temporal buffer assuming constant velocities (Hoffmann and Mortimer 1994), failing when pedestrians change speed unexpectedly. Post-encroachment time quantifies separation between successive zone occupancy (Ansariyar 2022) but only after conflict occurs. Time-to-collision with disturbance (TTCD) incorporates vehicle dynamics for better crash rate alignment (Xie *et al.* 2019), while curvilinear time-to-collision (CurvTTC) accounts for curved trajectories of both vehicles and pedestrians (Pu *et al.* 2025a), yet both require observable motion to compute risk. Despite these refinements, all surrogate safety metrics operate reactively, quantifying danger only after pedestrians enter conflict zones and providing insufficient lead time for proactive intervention.

Trajectory prediction methods forecast future pedestrian paths by extrapolating observed motion patterns under the assumption of continuous and smooth dynamics. Social LSTM introduced spatial pooling for interaction modeling (Alahi *et al.* 2016), enabling multi-agent trajectory forecasting but assuming smooth, continuous motion that fails when pedestrians make abrupt decisions. Attention mechanisms (Vemula *et al.* 2018) improved focus on relevant interactions yet remained unable to anticipate discrete intent switches preceding observable movement. Transformer architectures (Giuliari *et al.* 2021) leveraged long-range dependencies to refine predictions, while Trajectron++ achieved sub-0.5 m errors at 4.8 s horizons through heterogeneous graph representations (Salzmann *et al.* 2020), though both remained constrained by the assumption of continuous motion dynamics. This progression reveals a fundamental limitation: trajectory methods predict positions assuming motion continuity but fail when pedestrians suddenly yield, stop, or reverse direction based on discrete intent changes (Pu *et al.* 2026a).

Intent-from-motion classifiers attempt to infer crossing decisions from video cues. Early systems employed hand-crafted features including velocity thresholds and head orientation (Keller and Gavrila 2013), but struggled to capture complex behavioral patterns through manually designed rules. Modern approaches utilize end-to-end learning, with the Joint Attention in Autonomous Driving achieving 79% accuracy at 1 s horizons using CNN-LSTM architectures (Rasouli and Tsotsos 2018), demonstrating that deep learning can extract temporal patterns from video sequences. Fusing visual appearance cues with motion or trajectory information consistently improves prediction performance (Gadzicki *et al.* 2020), indicating that integrating complementary visual and kinematic features at higher processing stages enhances robustness under behavioral uncertainty. Despite these improvements, intent-from-motion classifiers exhibit pronounced sensitivity to regional traffic norms, with 25%–40% performance degradation under cross-regional transfer, suggesting that they internalize region-specific behavioral regularities rather than transferable social reasoning (Herman *et al.* 2021).

Existing systems fail to differentiate crossing behavior based on pedestrian demographics, particularly age, despite well-documented behavioral differences across populations (Teng *et al.* 2025). Children exhibit higher behavioral unpredictability and tend to accept shorter crossing gaps, while older adults generally walk more slowly and have longer decision times than working-age pedestrians (Yi *et al.* 2022, Osuret *et al.* 2024). Yet contemporary systems apply uniform safety margins regardless of age (Garikapati *et al.* 2024). This architectural blindness stems from data limitations: major datasets like Argoverse 2 (Wilson *et al.* 2023), nuScenes (Caesar *et al.* 2020), HDI dataset (Pu *et al.* 2025b), and Waymo (Sun *et al.* 2020) either lack age annotations entirely, preventing algorithms from learning demographic-specific patterns. The safety consequences of this omission are substantial.

## 2.2. Intent Inference in Autonomous Driving Architectures

Modern autonomous driving systems employ two dominant architectural paradigms for pedestrian interaction: modular perception-planning pipelines and end-to-end learning frameworks (Chen *et al.* 2023).

Modular architectures, exemplified by Waymo's self-driving system (Lindsey 2024) and the open-source Apollo platform (Feng and Zhang 2022), follow a sequential processing structure wherein perception modules detect and classify pedestrians, prediction modules forecast future trajectories, and planning modules compute collision-free paths (Liu and Diao 2024). In these systems, pedestrian intent is not represented as an explicit semantic variable but is instead embedded implicitly within distance-velocity thresholds that trigger emergency braking when time-to-collision falls below 1.5–2.0 seconds or spatial separation decreases to 10–15 meters (Bouhsain *et al.* 2020, Rasouli and Tsotsos 2020). Pedestrian yielding intent is not proactively inferred; instead, yielding decisions are abstracted into binary post-hoc labels derived from trajectory prediction modules. Consequently, behavioral cues—including hesitation patterns, eye contact, and body orientation—are systematically disregarded, as they fall outside the representational capacity of kinematic-based prediction models (Rouchitsas and Alm 2019, Li *et al.* 2024). As a result, intent is inferred only indirectly from observed motion statistics, rather than being modeled explicitly as a belief-driven decision variable, leaving a structural disconnect between perception outputs and decision-making logic.

End-to-end learning frameworks, such as Tesla's Full Self-Driving system (Suryana *et al.* 2025) and NVIDIA's PilotNet (Bojarski *et al.* 2020), bypass the modular autonomy pipeline by directly mapping sensor inputs to control actions using deep neural networks trained via behavioral cloning or reinforcement learning (Ranjan and Senthamilarasu 2020). While these approaches achieve strong performance on controlled benchmarks by exploiting correlations between visual patterns and driving actions, they exhibit critical limitations in safety-critical pedestrian interactions (Kuutti 2022). In particular, end-to-end policies generalize poorly under distributional shift, as models trained in one geographic or cultural context often degrade when deployed in environments with different traffic norms and pedestrian behaviors, reflecting reliance on dataset-specific correlations rather than transferable social reasoning about pedestrian intent (Bojarski *et al.* 2020, Camara *et al.* 2021). Moreover, the opaque nature of end-to-end decision-making prevents meaningful interpretation of control actions, making it impossible to assess whether a braking maneuver reflects correct intent inference or spurious reactive behavior, thereby undermining regulatory certification, failure analysis, and accident investigation (Lu *et al.* 2025b, Morgado *et al.* 2025). Fundamentally, both end-to-end and modular architectures fail to incorporate explicit intent reasoning, treating pedestrian behavior as a pattern recognition problem rather than a semantic inference task.

Both architectural paradigms share a common underlying assumption: that pedestrian yielding behavior can be reliably inferred from geometric configurations and kinematic observations. This assumption systematically fails in scenarios where yielding decisions are governed by discrete intent transitions rather than continuous motion dynamics (Munir *et al.* 2025). Pedestrians frequently initiate crossing or abort movement without kinematic precursors, producing instantaneous state transitions that violate the smoothness assumptions embedded in trajectory prediction models (Fu *et al.* 2024). Identical spatial configurations can lead to opposite pedestrian yielding decisions depending on social cues such as perceived driver attention or eye contact, rather than on physical constraints alone (Lyu *et al.* 2024). Because these decisions operate in semantic and social space and are shaped by traffic norms, risk perception, and Theory-of-Mind reasoning about driver intent, they cannot be reliably captured by architectures that rely solely on geometric configurations or learned feature representations (Scaliti *et al.* 2023, Wang *et al.* 2025).

In vehicle–pedestrian interactions, safe and efficient operation requires inferring pedestrian crossing and yielding intent before it manifests in observable motion patterns (Azarmi *et al.* 2025). Without explicit intent inference capabilities, autonomous vehicles often misclassify yielding pedestrians as non-yielding and initiate premature braking, even when pedestrians intend to yield right-of-way (Mirzabagheri *et al.* 2025). These false-positive interventions result in unnecessary deceleration in low-risk encounters, disrupting traffic flow, degrading intersection throughput, and increasing travel time. This limitation cannot be resolved by trajectory prediction or reactive learning, because pedestrian intent is governed by semantic cues and social norms rather than motion patterns alone (Lu *et al.* 2025a). As a result, current autonomous driving architectures lack the semantic intent reasoning layer necessary to support both safe and efficient vehicle–pedestrian interactions.

## 2.3. Research Gaps

The preceding review reveals three fundamental gaps constraining autonomous vehicle capabilities in pedestrian interaction scenarios: (1) inability to recognize pedestrian yielding intent, (2) lack of intent-aware decision-making causing unnecessary braking, and (3) absence of age-adaptive control strategies. First, current autonomous driving systems cannot recognize pedestrian yielding intent before observable motion. Motion-based approaches assume pedestrian behavior is continuous and governed by physical dynamics (Salzmann *et al.* 2020), yet yielding decisions are discrete and context-dependent, shaped by intent and social reasoning that precede kinematic changes (Wu *et al.* 2023). Existing systems cannot anticipate whether pedestrians will yield until movement occurs (Scaliti *et al.* 2023), forcing AVs to rely on geometric and kinematic cues while filtering out critical behavioral signals such as body orientation, gaze direction, and hesitation (Munir *et al.* 2025).

This inability directly causes unnecessary braking and inefficient traffic flow. When autonomous vehicles cannot determine pedestrian intent, they apply uniform braking responses to pedestrian proximity regardless of behavioral context (Sheikh and Peng 2025), producing frequent false-positive interventions for pedestrians who have already decided to yield. Both modular and end-to-end architectures lack semantic reasoning layers that translate geometric observations into interpretable intent states (Tian *et al.* 2024). Modular systems reduce intent to predefined distance–velocity thresholds that cannot capture nuanced behavioral cues (Munir *et al.* 2025), while end-to-end approaches learn direct sensor-to-control mappings that generalize poorly (Chen *et al.* 2023).

Contemporary autonomous driving architectures also lack age-adaptive control strategies. Existing systems apply uniform prediction and response logic despite well-documented evidence that children exhibit higher unpredictability and shorter gap acceptance, while seniors display reduced mobility and longer decision latencies (Yi *et al.* 2022, Li *et al.* 2023). Because major datasets lack demographic annotations, deployed models adopt homogeneous strategies that systematically underserve vulnerable populations (Pathiraja *et al.* 2024). Even when elevated risk is detected, current systems apply uniform braking without adapting to age-specific characteristics such as delayed reactions in seniors or erratic movements in children (Sheikh and Peng 2025), producing suboptimal safety margins for vulnerable populations. This reflects a fundamental architectural limitation: the absence of control frameworks integrating demographic awareness into real-time decision-making.

## 3. Methodology

This section presents the VLM-VPI framework, a closed-loop autonomous driving system that integrates multimodal large language models into CARLA for pedestrian intent prediction and demographic-adaptive vehicle control. The architecture comprises three key components: (1) a multimodal perception layer that captures visual and kinematic observations, (2) a vision-language reasoning layer that infers pedestrian intent and demographic categories through few-shot learning, and (3) a tiered safety control module that translates intent predictions into demographic-adaptive braking commands. The VLM-VPI employs a perception-reasoning-control framework operating in closed-loop mode within the CARLA simulation environment, as illustrated in Figure 1. The experimental testbed uses CARLA version 0.9.16 with the Town10HD high-definition urban map, which provides realistic intersection geometry, marked crosswalks, and traffic signal infrastructure representative of dense urban environments. The simulation operates in synchronous mode with a fixed time step of $\Delta t = 0.05$ s, corresponding to a 20 Hz control frequency, ensuring temporal consistency across sensor measurements, vehicle dynamics, and pedestrian motion states.

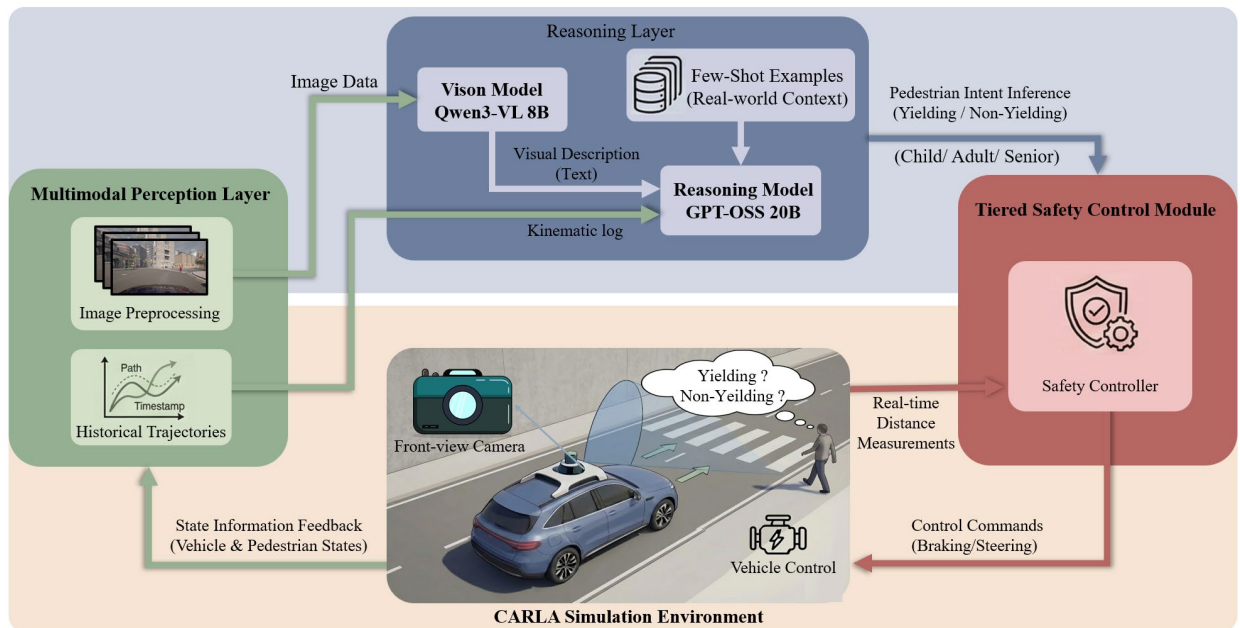

**Figure 1: The VLM-VPI framework architecture**

As illustrated in Figure 1, the system workflow operates through a multimodal perception layer that continuously monitors the CARLA simulation environment. The front-view camera captures images, while trajectory recorders maintain synchronized kinematic logs of vehicle and pedestrian states. When the vehicle enters an intersection zone and detects pedestrians, a single front-view image and corresponding kinematic data are transmitted to the reasoning layer. Within this layer, the vision model (Qwen3-VL 8B) converts the captured image into a structured textual scene description containing contextual, behavioral, and age-related visual cues. This description, together with kinematic logs and retrieved few-shot examples, is provided to the reasoning model (GPT-OSS 20B) for joint inference of pedestrian intent and age group (child, adult, or senior). The inferred semantic states are forwarded to the demographic-adaptive safety control module, which executes age-specific, distance-dependent braking strategies.

The tiered safety control module receives the pedestrian intent classification from the reasoning layer and continuously monitors real-time distance measurements from the perception layer. If the pedestrian is classified as non-yielding, the safety controller implements demographic-adaptive tiered braking strategies based on the pedestrian age category (child, adult, or senior) detected by the vision model during the visual description generation stage, mapping distance thresholds to appropriate deceleration levels ranging from 0.2g to 1.0g. Conversely, if the pedestrian is classified as yielding, the vehicle maintains its original CARLA autopilot control without intervention. Control commands are applied to the ego vehicle at 20 Hz frequency, and the resulting state changes feed back to the perception layer, forming a closed-loop system that continuously adapts to evolving traffic situations.

### 3.1. Multimodal Perception Layer

The multimodal perception layer comprises the sensing and initial processing components of the VLM-VPI framework, responsible for capturing visual and kinematic observations from the CARLA simulation environment. This layer operates through continuous monitoring, providing multimodal inputs to the reasoning layer for intent inference.

The multimodal perception layer comprises a front-view RGB camera capturing visual information at a resolution of $1280 \times 720$ pixels with a $120°$ field of view, positioned 1.2 m forward of the vehicle center at a height of 1.4 m above the ground with zero pitch angle, and a kinematic state recorder maintaining historical trajectories of vehicle and pedestrian states. At each simulation step $t$, the system records a comprehensive state vector $S_t$ capturing vehicle-pedestrian kinematics, including vehicle position $\left(x_{\text{veh}}^t, y_{\text{veh}}^t\right)$, vehicle velocity $\left(v_{\text{veh},x}^t, v_{\text{veh},y}^t\right)$, pedestrian position $\left(x_{\text{ped}}^t, y_{\text{ped}}^t\right)$, pedestrian velocity $\left(v_{\text{ped},x}^t, v_{\text{ped},y}^t\right)$, and bumper-to-pedestrian distance $d_t$. Visual attributes related to pedestrian posture, body scale, and apparent age are extracted separately by the vision model and incorporated into the subsequent semantic reasoning process. Pedestrian velocities are computed using finite-difference approximation as

$$v_{\text{ped},x}^t = \frac{x_{\text{ped}}^t - x_{\text{ped}}^{t-1}}{\Delta t}, v_{\text{ped},y}^t = \frac{y_{\text{ped}}^t - y_{\text{ped}}^{t-1}}{\Delta t} \tag{1}$$

The vehicle front reference point $b_t$ is defined based on the vehicle position, and the Euclidean distance $d_t = \left\| b_t - p_{\text{ped}}^t \right\|$ serves as the primary safety metric.

The system employs event-triggered activation to balance computational efficiency with early semantic intent inference, ensuring that both behavioral and demographic cues are captured before critical interaction stages. LLM inference activates when three conditions are simultaneously satisfied:

$$\text{Trigger } = (d_t < d_{\text{thresh}}) \wedge \left( \text{IsJunction}\left(p_{\text{veh}}^t\right)\right) \wedge (\neg \text{ HasTriggered }) \tag{2}$$

where $d_{\text{thresh}} = 15$ m defines the activation distance threshold. IsJunction $\left(p_{\text{veh}}^t\right)$ evaluates whether the vehicle occupies a junction region based on CARLA waypoint topology, and HasTriggered prevents redundant invocations within a single interaction episode. Upon trigger activation, the system suspends autopilot control, captures the current front-view camera frame as a PNG file, and exports the accumulated trajectory buffer containing the full historical kinematic records of the ongoing interaction episode. The multimodal input vector $\mathcal{I}_t$ integrates visual and temporal information, defined as $\mathcal{I}_t = \{I_t, \mathcal{T}_{hist}\}$. Here, $I_t$ denotes the front-view image at trigger instant $t$, and $\mathcal{T}_{hist} = \{s_{t_0}, s_{t_0+1}, \dots, s_t\}$ represents the complete kinematic trajectory from the episode start time $t_0$ to the trigger instant. The trajectory is formatted as a temporally ordered JSON array, preserving the full motion history of both the vehicle and the pedestrian. The front-view image is processed by the vision model (Qwen3-VL 8B) to generate a textual description of the visual scene, focusing on pedestrian count and locations, body orientations relative to the roadway, apparent gaze directions, walking or standing postures, proximity to crosswalk boundaries, and the presence of distractions. The kinematic trajectory is represented as a temporally ordered sequence $\mathcal{T}_{hist}^{(t)} = \{s_\tau\}_{\tau=t_0}^{t}$, where each state vector is defined as

$$\mathbf{s}_\tau = \left(x_{veh}^\tau, y_{veh}^\tau, v_{veh,x}^\tau, v_{veh,y}^\tau, x_{ped}^\tau, y_{ped}^\tau, v_{ped,x}^\tau, v_{ped,y}^\tau, d_\tau\right)^\top.$$

This state vector encodes the vehicle and pedestrian positions and velocities in the CARLA world frame, together with the bumper-to-pedestrian distance. The resulting trajectory sequence is provided to the reasoning model as the {kinematic_data} input.

### 3.2. Reasoning Layer

The reasoning module incorporates domain-specific behavioral knowledge through few-shot prompting grounded in real-world vehicle pedestrian interactions. Instead of relying solely on the model's general prior knowledge, the system provides a small set of curated reference cases with verified intent labels and structured reasoning annotations. These six exemplars, stratified by age and intent, function as high-dimensional behavioral priors that calibrate the LLM's reasoning to cross-domain pedestrian dynamics, enabling the model to generalize from structured real-world interaction patterns to diverse simulated scenarios.

The few-shot exemplar set is constructed from annotated samples extracted from the Pedestrian Intention Estimation (PIE) dataset (Rasouli *et al.* 2019). PIE contains over six hours of high definition on board video collected on Toronto streets at 30 FPS, synchronized with ego vehicle kinematic measurements from OBD sensors, including vehicle speed, GPS coordinates, and heading direction. In this study, the exemplar set contains a fixed set of representative cases organized by demographic category and intent class. It includes three yielding cases and three non-yielding cases, with one exemplar for each demographic intent combination, namely Child Yielding, Adult Yielding, Senior Yielding, Child Non Yielding, Adult Non Yielding, and Senior Non Yielding. During inference, all six exemplars are consistently included in the prompt to provide stable behavioral priors rather than being dynamically selected. Figure 2 demonstrates this comprehensive data structure through a representative "Adult Yielding" case, where the exemplar integrates three key components: the ego-centric visual context (Figure 2 (a)), the raw kinematic state trajectory (Figure 2 (b)), and the corresponding structured reasoning annotation that provides expert-level interpretation (Figure 2 (c)). This multi-modal representation enables the system to ground its predictions in concrete, annotated real-world scenarios.

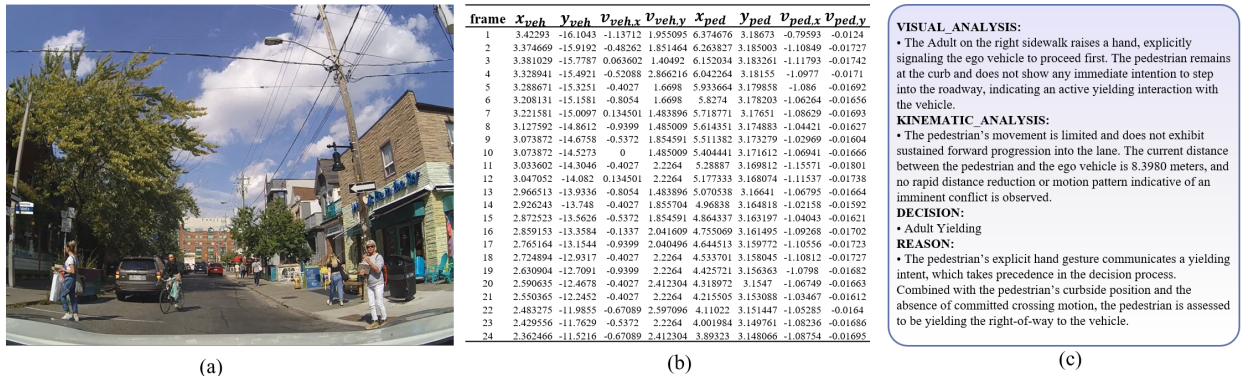

**Figure 2: A real-world adult yielding exemplar: (a) ego-centric vision, (b) kinematic state log, and (c) structured reasoning**

Each exemplar is formatted as a user-assistant dialogue pair, with the user message containing the vision model's textual scene description and JSON-formatted kinematic log prefaced with "Sample Input (Real-World Reference):", and the assistant message containing the ground-truth output following standardized four-section format including visual analysis, kinematic analysis, binary decision, and explicit reasoning.

The vision model (Qwen3-VL 8B) processes the front-view camera image to generate a structured textual description. The reasoning model (GPT-OSS 20B) then performs intent inference through a structured prompt template, illustrated in Figure 3. The template specifies four sections: a role declaration that establishes the operational context; a task instruction that defines the inference objective; three input placeholders ({visual_description}, {kinematic_data}, {few_shot_examples}) populated at runtime; and an output format schema that constrains the model to produce four mandatory fields. The VISUAL_ANALYSIS field examines spatial cues from the vision description, including pedestrian position relative to the crosswalk, body orientation with respect to the roadway, apparent gaze direction, and limb positioning. The KINEMATIC_ANALYSIS field evaluates quantitative features in the trajectory data, including velocity magnitude trends, velocity direction consistency, vehicle–pedestrian distance closure rate, and sudden motion pattern changes. The DECISION field requires commitment to one of two mutually exclusive classifications, and the REASON field demands explicit causal argumentation linking observations to the classification. This template-driven decomposition ensures that every yielding inference is accompanied by a traceable evidence chain, supporting both interpretability and post-hoc verification.

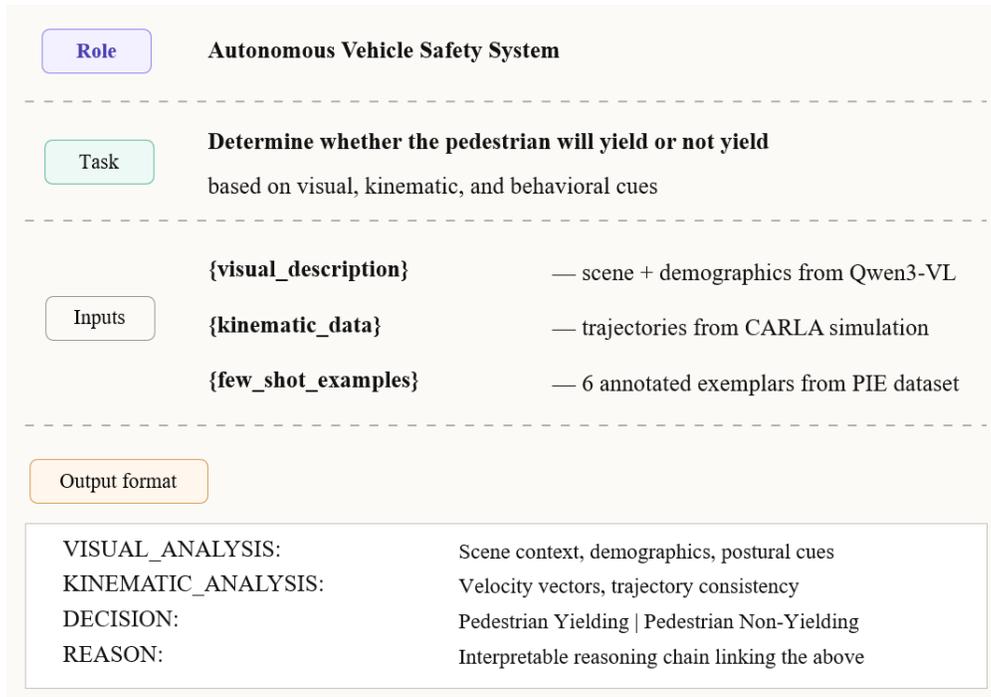

**Figure 3: Prompt template for pedestrian yielding inference**

Upon receiving the reasoning model output, the system extracts the jointly inferred pedestrian intent classification and demographic category from the structured response. The inferred intent determines whether the vehicle transitions to emergency control or continues under autopilot, while the detected demographic category (Child, Adult, or Senior) is passed to the tiered safety control module to activate the corresponding demographic-specific safety parameters. To ensure safe operation, conservative fallback logic is applied in cases of ambiguous or invalid outputs. The complete reasoning result is logged for post hoc interpretability and failure-mode analysis.

Table 1 summarizes the vision-language model configuration and few-shot learning parameters. The vision model operates with 8 billion parameters generating textual scene descriptions with maximum output length of 256 tokens. The reasoning model employs a context window of 128,000 tokens to accommodate comprehensive prompt structure including system instructions, multiple few-shot exemplars, vision descriptions, and kinematic trajectory data.

**Table 1 Vision-language model configuration and few-shot learning parameters**

| Category | Parameter | Configuration/Value |
|---|---|---|
| Vision Model | Architecture | Qwen3-VL 8B |
| | Model Size | 8 billion parameters |
| | Input Resolution | $1280 \times 720$ pixels |
| | Output Length | 256 tokens |
| Reasoning Model | Architecture | GPT-OSS 20B |
| | Model Size | 20 billion parameters |
| | Context Window | 128,000 tokens |
| | Max Output Tokens | 500 |
| Few-Shot Database | Data Source | PIE Dataset (Toronto, Canada) |
| | Total Exemplars | 6 annotated interactions |
| | Yielding Cases | 3 exemplars |
| | Non-Yielding Cases | 3 exemplars |
| | Demographic Stratification | Child (2 exemplars), Adult (2 exemplars), Senior (2 exemplars). (2 exemplars: 1 yielding + 1 non-yielding) |
| Prompt Structure | System Prompt | Role definition + output format |
| | Few-Shot Section | User-assistant dialogue pairs |
| | Current Query | Vision description + kinematic JSON |
| | Total Token Budget | ~8,000-12,000 tokens |

### 3.3. Tiered Safety Control Module

The tiered safety control module translates the jointly inferred pedestrian intent and age category from the reasoning layer into precise longitudinal control commands that ensure collision avoidance while accommodating demographic-specific collision risks through adaptive safety margins. This module operates continuously at 20 Hz throughout the simulation, executing control strategies that scale braking intensity based on both collision imminence and pedestrian demographic characteristics. The system maintains a persistent control mode state that governs whether the vehicle operates under standard autopilot or manual emergency control. The control mode is formally defined as:

$$\text{Mode}_t = \begin{cases} \text{AUTOPILOT}, & \text{if Decision} = \text{Yielding} \\ \text{EMERGENCY}, & \text{if Decision} = \text{Non-Yielding} \end{cases} \tag{3}$$

In AUTOPILOT mode, the CARLA Traffic Manager provides longitudinal and lateral control commands based on its built-in path planning algorithms, with the vehicle maintaining its lane and respecting speed limits adjusted to 70% of nominal values. In EMERGENCY mode, the Traffic Manager's control authority is revoked and the system applies manual control commands computed by the demographic-adaptive tiered braking algorithm. The mode transition from AUTOPILOT to EMERGENCY occurs immediately upon receiving a non-yielding classification from the reasoning model.

A critical innovation of the proposed control architecture is the integration of demographic-specific safety factors that modify distance thresholds and braking intensities based on pedestrian age category. Traffic safety research indicates systematic differences in pedestrian collision risk across demographic groups, which motivate demographic-adaptive safety margins in vehicle control. Child pedestrians exhibit higher behavioral variability and reduced hazard perception, increasing the likelihood of abrupt or unpredictable crossing actions (Kendi and Johnston 2023). Senior pedestrians, by contrast, experience age-related mobility decline, leading to slower walking speeds and longer exposure times within the conflict zone (Asher *et al.* 2012). Adults generally demonstrate more stable and predictable crossing behavior and are therefore treated as the baseline group. These documented differences justify the use of larger safety buffers for children and seniors to ensure sufficient reaction time and conservative braking under uncertainty (Ishaque and Noland 2008). To accommodate these differential risk profiles, the control system implements demographic-specific safety margin multipliers:

$$\alpha_{\text{demo}} = \begin{cases} 1.4, & \text{Child (age} < 18) \\ 1.0, & \text{Adult (age 18-65)} \\ 1.2, & \text{Senior (age} > 65) \end{cases} \tag{4}$$

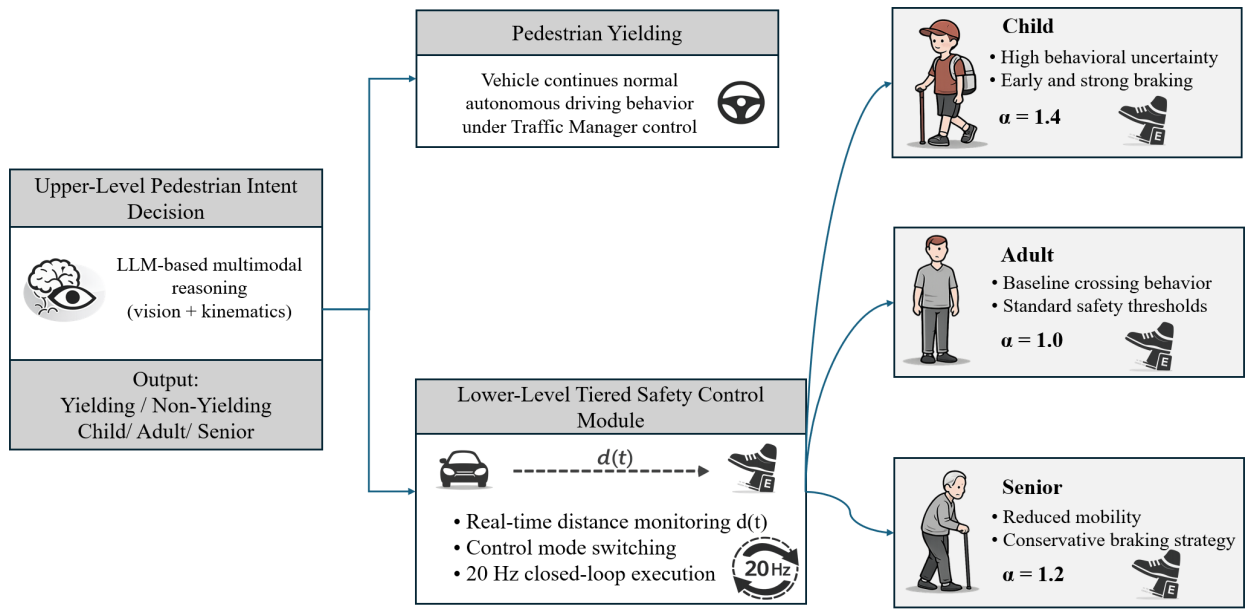

**Figure 4: Demographic-Adaptive Tiered Safety Control Architecture**

These multipliers are applied to baseline distance thresholds defining braking tier transitions, effectively creating demographic-specific protective zones. The child safety factor of 1.4 and senior factor of 1.2 were calibrated based on documented collision risk profiles and validated through sensitivity analysis (Table 6). To prevent control mode oscillation while accounting for demographic-specific risks, the system incorporates a multi-criteria resume mechanism that evaluates safety restoration through spatial, temporal, and behavioral dimensions rather than relying solely on longitudinal distance thresholds.

The trigger condition activating emergency control remains distance-based with demographic scaling:

$$d_{\text{trigger}} = 9.35 \times \alpha_{\text{demo}} \tag{5}$$

Adult pedestrians maintain the baseline threshold of 9.35 m. This value is calibrated based on minimum AEB activation distances observed in empirical evaluations of 2025 model year vehicles, providing a real-world reference point that is adapted to CARLA's physics engine through preliminary testing (American Automobile 2025). Intent inference is intentionally triggered at a distance greater than this baseline braking threshold. This design ensures that pedestrian intent is predicted before the vehicle enters the empirically defined critical braking zone, providing sufficient time for the tiered safety control strategies to select and execute an appropriate braking strategy without relying on last-moment emergency intervention.

The core of the control strategies is a tiered, piecewise-constant braking policy that maps the instantaneous bumper-to-pedestrian distance $d_t$ to a commanded longitudinal deceleration. The policy is designed to emulate human defensive driving behavior, whereby drivers adopt progressively stronger braking responses as collision imminence increases. Rather than applying a single uniform braking strategy, the proposed controller adapts its distance thresholds according to pedestrian demographic, expanding protective buffers for populations exhibiting higher behavioral uncertainty or reduced mobility. Formally, the control law can be expressed as a demographic-parameterized mapping $a_{\text{cmd}}(d_t; \alpha_{\text{demo}})$, where $a_{\text{cmd}}$ denotes the commanded longitudinal deceleration. The resulting control action is piecewise constant, with discrete deceleration levels corresponding to mild, moderate, heavy, and emergency braking regimes.

Table 2 summarizes the demographic-adaptive braking distance thresholds for each tier. Compared to the adult baseline, both child and senior categories receive expanded thresholds. The four-tier structure employs deceleration levels of 0.2 g, 0.4 g, 0.7 g, and 1.0 g, calibrated based on human comfort thresholds and vehicle stability constraints. The mildest tier (0.2 g) remains below the threshold where passengers typically perceive noticeable deceleration, the moderate tier (0.4 g) represents assertive braking that communicates stopping intent, the heavy tier (0.7 g) approximates a near-limit deceleration level under favorable dry-road conditions, and the emergency tier (1.0 g) represents an upper-bound theoretical deceleration intended to approximate extreme emergency braking under ideal tire–road friction conditions.

**Table 2 Demographic-adaptive braking distance thresholds**

| Braking Tier | Deceleration | Child Thresholds (m) | Adult Thresholds (m) | Senior Thresholds (m) |
|---|---|---|---|---|
| Mild (Tier 1) | 0.2 g | 19.64 - 26.18 | 14.03 - 18.70 | 16.83 - 22.44 |
| Moderate (Tier 2) | 0.4 g | 13.09 - 19.64 | 9.35 - 14.03 | 11.22 - 16.83 |
| Heavy (Tier 3) | 0.7 g | 6.55 - 13.09 | 4.68 - 9.35 | 5.61 - 11.22 |
| Emergency (Tier 4) | 1.0 g | < 6.55 | < 4.68 | < 5.61 |

The implementation in CARLA maps the commanded acceleration $a_{\text{cmd}}$ to the simulator's normalized brake input $\beta \in [0,1]$ using $\beta = |a_{\text{cmd}}|/g$, corresponding to brake inputs of 0.2, 0.4, 0.7, and 1.0 for the four tiers respectively. At each control cycle, the system computes the current bumper-to-pedestrian distance, evaluates the piecewise braking law to determine the appropriate brake intensity, and constructs a CARLA VehicleControl object with throttle set to zero, steering set to the previous value to maintain lateral trajectory, and brake set to the computed intensity.

However, the resume condition—which transitions the system from EMERGENCY mode back to AUTOPILOT—employs a fundamentally different logic structure. Recognizing that purely longitudinal braking cannot increase vehicle-pedestrian distance without lateral displacement or pedestrian motion, the system implements a three-dimensional safety assessment:

$$\text{Resume} = \text{EMERGENCY} \wedge \left( C_{\text{spatial}} \vee C_{\text{temporal}} \vee C_{\text{behavioral}} \right) \tag{6}$$

where the three constituent conditions are defined as follows:

Spatial Clearance Condition ( $C_{\text{spatial}}$ ) evaluates whether the pedestrian has laterally exited the vehicle's trajectory corridor:

$$C_{\text{spatial}} = \left( d_{\perp,t} > d_{\text{clearance}} \right) \tag{7}$$

where $d_{\perp,t} = \min_s \left\| p_{\text{ped}}^t - p_{\text{path}}(s) \right\|$ represents the minimum perpendicular distance from the pedestrian position $p_{\text{ped}}^t$ to any point $s$ along the vehicle's planned trajectory $p_{\text{path}}(s)$, and the clearance threshold is demographically scaled as $d_{\text{clearance}} = 3.5 \times \alpha_{\text{demo}}$ meters. This yields clearance distances of 4.9 m for children, 3.5 m for adults, and 4.2 m for seniors. The spatial condition captures scenarios where pedestrians complete crossing maneuvers or retreat to sidewalks, creating unobstructed corridors for vehicle progression.

Temporal Safety Window Condition ( $C_{\text{temporal}}$ ) requires sustained vehicle immobilization with maintained safe separation:

$$C_{\text{temporal}} = \left( v_{\text{veh},t} < \frac{0.1\ \text{m}}{\text{s}} \right) \wedge \left( t - t_{\text{stop}} > T_{\text{wait}} \right) \wedge \left( d_t > d_{\text{min\_safe}} \right) \tag{8}$$

where $v_{\text{veh},t}$ denotes instantaneous vehicle speed, $t_{\text{stop}}$ marks the timestamp when speed first fell below $0.1\ \text{m/s}$, $T_{\text{wait}} = 2.0\ \text{s}$ defines the minimum observation window, and $d_{\text{min\_safe}} = 5.0 \times \alpha_{\text{demo}}$ meters specifies demographically adjusted standoff distance (7.0 m for children, 5.0 m for adults, 6.0 m for seniors). This condition addresses situations where pedestrians remain stationary near the roadway but maintain sufficient distance from the stopped vehicle, allowing safe resumption after a prudent waiting period that ensures pedestrian intentions have stabilized. In CARLA simulation, pedestrian behaviors follow the default walker controller. Non-yielding pedestrians typically complete their crossing maneuver after the initial conflict detection, thereby satisfying the spatial clearance condition ($C_{\text{spatial}}$) as they exit the vehicle's trajectory corridor. In cases where pedestrians remain stationary, the temporal safety window condition ($C_{\text{temporal}}$) ensures safe resumption after the vehicle has come to a complete stop and maintained sufficient separation for the observation period.

Behavioral Indication Condition ($C_{\text{behavioral}}$) detects pedestrian movement patterns indicating active withdrawal from the conflict zone:

$$C_{\text{behavioral}} = \langle \text{v}_{\text{ped},t}, \text{n}_{\text{away}} \rangle > v_{\text{threshold}} \tag{9}$$

where $\text{v}_{\text{ped},t}$ is the pedestrian velocity vector, $\text{n}_{\text{away}}$ is the unit normal vector pointing away from the vehicle's planned path toward the nearest safe zone (typically the sidewalk), and $v_{\text{threshold}} = 0.5\ \text{m/s}$ represents the minimum speed indicating purposeful retreat. The inner product $\langle \cdot, \cdot \rangle$ projects pedestrian velocity onto the escape direction, filtering lateral motion components that do not contribute to effective risk reduction.

The disjunctive logic structure ($\vee$) ensures that satisfying any single condition is sufficient to authorize mode transition, providing multiple independent pathways for conflict resolution while maintaining conservative safety standards. This design accommodates diverse pedestrian behaviors: crossing completion (spatial), stationary yielding (temporal), or active retreat (behavioral). When any condition is met, the system re-enables CARLA Traffic Manager autopilot control and resets the trigger state flag HasTriggered to permit future reasoning model invocations in subsequent interactions.

The control strategies operate in closed-loop mode, continuously updating control commands based on real-time state feedback. At each simulation tick, the system queries the current vehicle and pedestrian states from CARLA, computes the updated bumper-to-pedestrian distance, evaluates the control mode and demographic-adaptive braking law, and applies the resulting control command. This closed-loop architecture enables the controller to respond effectively to unexpected changes in pedestrian behavior and interaction dynamics.

The complete closed-loop system dynamics can be represented as:

$$S_{t+1} = f(S_t, u_t, \text{Mode}_t, \alpha_{\text{demo}}, \zeta_t, w_t) \tag{10}$$

where $S_{t+1}$ denotes the state vector at the next time step, $f(\cdot)$ represents the vehicle dynamics function implemented by CARLA's physics engine, $u_t$ is the control input, Mode $_t$ is the current control mode, $\alpha_{\text{demo}}$ is the demographic safety factor, and $w_t$ represents external disturbances including pedestrian motion and environmental factors. The variable $\zeta_t$ denotes the semantic pedestrian intent state inferred by the vision-language reasoning layer at time $t$, incorporating both yielding classification and demographic category. By explicitly integrating semantic intent inference with demographic-adaptive control and physical vehicle dynamics, the proposed architecture forms a unified closed-loop autonomous driving system capable of handling ambiguous pedestrian behaviors while providing consistent and risk-sensitive safety protection for vulnerable road users.

## 4. Results and Discussion

The evaluation uses three distinct datasets generated from CARLA Town10HD simulations, each designed for specific experimental objectives: Dataset 1 (112 scenarios) for intent classification performance evaluation (Section 4.1), containing diverse behavioral patterns across varying ambiguity levels; Dataset 2 (180 scenarios, stratified by demographics: 60 children, 60 adults, 60 seniors) for demographic-adaptive control validation (Section 4.2), with balanced intent distributions within each age group; Dataset 3 (200 scenarios with randomized vehicle speeds 25-35 km/h and pedestrian speeds 2-4 m/s) for safety and efficiency assessment including TTC dynamics and false alarm rates (Sections 4.4 and 4.5). These datasets are independent and non-overlapping to ensure unbiased evaluation across different performance dimensions.

### 4.1. Pedestrian Crossing Intent Classification Performance

The primary objective of the VLM-VPI framework is to accurately infer pedestrian crossing intentions from multimodal observations while evaluating the contribution of real-world behavioral priors beyond pre-trained model knowledge. Accordingly, classification performance is assessed across 112 test scenarios in CARLA Town10HD using a comprehensive ablation framework that isolates the effects of visual perception, reasoning model integration, and authentic few-shot exemplars. Five system configurations with progressively enhanced capabilities are evaluated. The rule-based baseline was heuristically constructed using threshold values informed by pedestrian automatic emergency braking (PAEB) response distance observations reported in the American Automobile Association evaluation (American Automobile 2025). In this baseline, pedestrians were considered non-yielding when the bumper distance was below 9.35 m and their velocity magnitude toward the vehicle exceeded 0.8 m/s. The vision-only LLM configuration employs the Qwen3-VL 8B model to generate scene descriptions processed through rule-based logic, isolating pure visual perception. The kinematics-only LLM configuration provides the GPT-OSS 20B reasoning model with JSON-formatted trajectory data only, isolating temporal motion pattern analysis. The zero-shot LLM configuration integrates full multimodal inputs without real-world exemplars, relying solely on pre-trained knowledge. The real-world few-shot LLM corresponds to the proposed VLM-VPI framework and incorporates six annotated exemplars from the PIE dataset.

To enable supervised learning baseline comparisons, we first manually verified the ground-truth intent labels and trajectory quality for all 112 scenarios to ensure annotation reliability and data integrity. Based on the verified dataset, the scenarios were then partitioned into training (78 scenarios, 70%) and testing sets (34 scenarios, 30%) using stratified random sampling to maintain balanced representation across demographic categories and intent classes. For machine learning baseline comparisons, we implement three supervised learning methods trained on the verified trajectory data consisting of 10 fields per timestep: frame index, vehicle position, vehicle velocity, pedestrian position, pedestrian velocity, and bumper-to-pedestrian distance. SVM classifier (Yang *et al.* 2015) with radial basis function kernel (gamma=0.1, C=1.0) uses 12-dimensional aggregated statistical features including mean and standard deviation of pedestrian velocity and distance, speed extrema, relative velocity, velocity direction consistency, and distance change rate. LSTM-based trajectory classifier (Ji *et al.* 2020) employs bidirectional architecture (2 layers, 128 hidden units per direction, dropout rate 0.3) processing 40-step sequences with 9-dimensional feature vectors per timestep, trained using Adam optimizer (learning rate 0.001, batch size 16, 100 epochs with early stopping). CAPformer (Lorenzo *et al.* 2021) implements transformer encoder (4 layers, 8 attention heads, 256 embedding dimensions, feedforward dimension 1024) processing the same sequences with multi-head self-attention mechanisms and two-layer MLP classification head (hidden dimension 128), trained using identical settings as LSTM.

**Table 3** presents comprehensive classification performance across all eight configurations, enabling direct quantification of each component's marginal contribution. The table reports four key metrics: Accuracy measures overall classification performance across both yielding and non-yielding cases; Recall (Non-Yield) quantifies the proportion of dangerous non-yielding pedestrians correctly identified, representing the most critical safety metric; False negative rate indicates the percentage of non-yielding pedestrians incorrectly classified as yielding—the most dangerous error type leading directly to collision risk; and marginal contribution shows the incremental accuracy improvement attributable to each added component. The proposed VLM-VPI achieves 92.3% accuracy with 94.1% recall and 5.9% false negative rate on non-yielding cases.

**Table 3 Intent classification performance across 112 test scenarios**

| Method | Accuracy | Recall (Non-Yield) | False Negative Rate | Marginal Contribution |
|---|---|---|---|---|
| Rule-based Baseline | 78.4% | 81.3% | 18.7% | Baseline |
| SVM Classifier | 73.5% | 75.0% | 25.0% | -4.9% |
| LSTM Trajectory Classifier | 79.4% | 81.8% | 18.2% | +1.0% |
| CAPformer | 82.4% | 84.6% | 15.4% | +4.0% |
| Vision-only LLM | 82.8% | 84.0% | 16.0% | +4.4% (vision) |
| Kinematics-only LLM | 83.9% | 85.3% | 14.7% | +5.5% (kinematics) |
| Zero-shot LLM | 88.4% | 89.8% | 10.2% | +10.0% (reasoning integration) |
| Real-world Few-shot LLM | 92.3% | 94.1% | 5.9% | +13.9% (real-world priors) |

The progression reveals critical findings regarding the comparative advantages of different methodological paradigms. Multimodal reasoning integration through zero-shot LLM (88.4%) provides substantial improvement over all baseline methods, exceeding individual modalities (vision: 82.8%, kinematics: 83.9%), supervised learning approaches (SVM: 73.5%, LSTM: 79.4%, CAPformer: 82.4%), and rule-based control (78.4%). Notably, the zero-shot LLM configuration achieves this performance without any task-specific fine-tuning on the CARLA dataset, whereas the supervised baselines still underperform by 9.9 percentage points (CAPformer) to 18.8 percentage points (SVM). Real-world few-shot LLM achieves the highest overall performance (92.3%), contributing an additional 3.9 percentage points beyond zero-shot and reducing false negative rate from 10.2% to 5.9%.

The supervised learning baselines reveal fundamental limitations of pure data-driven approaches in this domain. SVM (73.5%) underperforms even the rule-based baseline (78.4%), indicating that hand-crafted features from kinematic data provide insufficient discriminative power without temporal modeling or semantic understanding. LSTM (79.4%) improves over SVM (73.5%) by 5.9 percentage points through capturing temporal dependencies in motion patterns, achieving performance comparable to rule-based methods (78.4%) and demonstrating that recurrent architectures can learn meaningful trajectory representations. CAPformer (82.4%) represents the current state-of-the-art in trajectory-based intent prediction, leveraging transformer self-attention to model complex temporal patterns and approaching the performance of vision-only LLM (82.8%). However, all three supervised methods trained on 78 scenarios underperform real-world few-shot LLM (92.3%) that uses only six exemplars, highlighting a critical insight: domain-specific behavioral priors from authentic human behavior provide more effective learning signals than larger quantities of simulation-generated training data, establishing few-shot grounding as a sample-efficient solution for safety-critical deployment where extensive labeled datasets are costly or unavailable.

Performance stratification across scenario complexity reveals where each component provides greatest value. Table 4 presents accuracy metrics for clear intent scenarios (n=63) with unambiguous crossing intentions, moderate ambiguity scenarios (n=35) with mixed or time-varying signals including hesitation behaviors, and high ambiguity scenarios (n=14) featuring contradictory cues including false starts or direction reversals.

**Table 4 Classification performance stratified by scenario complexity**

| Scenario Category | Rule-based Baseline | Vision-only LLM | Kinematics-only LLM | Zero-shot LLM | Real-world Few-shot LLM | Improvement |
|---|---|---|---|---|---|---|
| Clear Intent ( n = 63 ) | 87.3% | 90.5% | 91.3% | 96.8% | 98.4% | +1.6% |
| Moderate Ambiguity ( n = 35 ) | 68.6% | 77.1% | 80.0% | 85.7% | 91.4% | +5.7% |
| High Ambiguity ( $n = 14$ ) | 42.9% | 50.0% | 57.1% | 64.3% | 71.4% | +7.1% |

*Improvement = Real-world few-shot accuracy minus Zero-shot accuracy.

The real-world few-shot LLM configuration provides progressively larger accuracy improvements over the zero-shot LLM baseline as scenario complexity increases, yielding gains of 1.6 percentage points for Clear Intent, 5.7 percentage points for Moderate Ambiguity, and 7.1 percentage points for High Ambiguity. In Clear Intent scenarios, the zero-shot LLM already achieves high accuracy (96.8%) because behavioral cues are sufficiently unambiguous for general pre-trained knowledge to support reliable inference. In contrast, Moderate and High Ambiguity scenarios require disambiguation of hesitation patterns and false starts, where authentic real-world exemplars supplied to the real-world few-shot LLM provide domain-specific behavioral references that cannot be replicated by generic pre-training alone. The observed improvement gradient indicates that domain-specific behavioral priors become increasingly valuable in interaction contexts where general reasoning proves insufficient. The kinematics-only LLM configuration demonstrates notably strong performance in High Ambiguity scenarios (57.1%), exceeding the vision-only LLM by 7.1 percentage points. This suggests that temporal velocity pattern analysis provides more robust intent cues under high uncertainty than static spatial features, as oscillatory motion patterns with measurable variance capture hesitation dynamics that single-frame posture information cannot represent.

Overall, performance stratification shows that VLM-VPI, implemented as the real-world few-shot LLM configuration, achieves near-perfect accuracy (98.4%) under Clear Intent conditions, maintains strong performance (91.4%) in moderately ambiguous scenarios through multimodal reasoning, and remains competitive (71.4%) in highly ambiguous cases characterized by intrinsically unpredictable pedestrian behavior. The consistent accuracy gains from the zero-shot LLM to the real-world few-shot LLM across all complexity levels highlight the importance of authentic behavioral grounding for safety-critical pedestrian intent prediction.

## 4.2. Demographic-Adaptive Control Validation

The tiered safety control strategies incorporate demographic-specific safety factors ( $\alpha_{\text{demo}} = \{1.4, 1.0, 1.2\}$ for children, adults, and seniors) that expand braking distance thresholds for vulnerable populations. This section validates whether demographic adaptive control provides measurable safety improvements compared to uniform control approaches that apply identical thresholds regardless of pedestrian age.

To isolate the contribution of demographic-adaptive control, we compare three controller configurations operating within identical simulation environments. The Rule-based Baseline represents conventional threshold-based emergency braking without LLM reasoning integration, applying uniform 9.35 m trigger threshold and single-tier 1.0g deceleration when distance and velocity criteria are simultaneously satisfied, serving as the current industry practice benchmark. The Uniform Control configuration integrates the complete VLM-VPI framework with tiered safety control but applies fixed distance thresholds across all pedestrian demographics ($\alpha$_demo = 1.0 for all age categories) and tiered braking distances following Table 2 adult baseline values regardless of detected pedestrian age. The Demographic-Adaptive Control configuration represents the proposed complete system implementing demographic-specific safety factors, scaling thresholds according to $\alpha$_demo values of 1.4 for children, 1.0 for adults, and 1.2 for seniors, expanding both trigger/resume thresholds and tiered braking boundaries as specified in Table 2. Both the Uniform Control and Demographic-Adaptive Control configurations receive identical pedestrian intent classifications from the upper-level reasoning module (Real-world Few-shot LLM), ensuring performance differences reflect purely control-layer threshold design rather than intent prediction accuracy variations.

The test dataset comprises 180 scenarios stratified by ground-truth pedestrian demographics: 60 child scenarios (age <18), 60 senior scenarios (age >65), and 60 adult scenarios (age 18-65). Each demographic subset includes 40 non-yielding and 20 yielding cases to maintain balanced intent distribution. Pedestrian behaviors follow CARLA's default walker controller without modification, as demographic adaptation occurs entirely within the vehicle control system through threshold scaling. The vision model (Qwen3-VL 8B) detects pedestrian age categories during scene description generation, providing demographic labels that activate appropriate $\alpha$_demo multipliers in the control module.

Time-to-Collision (TTC) is adopted as a surrogate safety indicator to quantify the temporal margin to potential collision in vehicle-pedestrian interactions. TTC represents the remaining time until a collision would occur if the current relative motion between the vehicle and the pedestrian were maintained. TTC is continuously evaluated over the closed-loop interaction horizon. At each simulation time step $t$, TTC is computed based on the instantaneous bumper-to-pedestrian distance $d_t$ and its temporal rate of change. The distance derivative $\frac{dd_t}{dt}$ characterizes relative motion, where negative values indicate approaching trajectories and non-negative values indicate constant separation or divergence. Accordingly, TTC is defined as

$$TTC_t = \begin{cases} \frac{d_t}{-\frac{dd_t}{dt}}, & \text{if } \frac{dd_t}{dt} < 0 \\ \text{undefined}, & \text{otherwise.} \end{cases} \quad (11)$$

This formulation restricts TTC computation to converging trajectories with collision potential. When the relative distance is constant or increasing, TTC is considered undefined and excluded from safety-critical analysis. For each interaction episode, safety severity is summarized by the minimum TTC over the closed-loop horizon, denoted as $\min_t TTC_t$, which captures the most critical moment of the vehicle-pedestrian interaction. Based on this measure, a conflict event is defined as an episode in which $\min_t TTC_t < 2.0$ s. This threshold ensures that the evaluation captures near-miss events in which the available reaction window falls below typical human–robot interaction safety margins (Zhang *et al.* 2025).

Table 5 presents conflict rate, false negative rate, and minimum TTC statistics stratified by pedestrian demographic category. The Demographic-Adaptive Control configuration substantially reduces conflict occurrence for vulnerable populations compared to Uniform Control. Child pedestrian scenarios achieve mean minimum TTC of 4.3 s under adaptive control versus 3.6 s under uniform control, representing 19% increase in temporal safety buffer. Conflict event frequency decreases by 60% for children (4 versus 10 events) and 54.5% for seniors (5 versus 11 events) when comparing adaptive to uniform control. Adult scenarios demonstrate equivalent performance between uniform and adaptive controllers (both maintain 4.5 s mean TTC with 2 conflict events), confirming that demographic adaptation does not compromise baseline safety while providing targeted protection for vulnerable groups. False negative rate remains identical across uniform and adaptive controllers (7.5% for children and seniors, 2.5% for adults), validating that expanded safety margins do not alter intent classification accuracy from the reasoning module. The false negative cases observed in the child and senior categories (three cases each) reflect ambiguous scenarios where pedestrians exhibited contradictory behavioral cues (initial stationary posture followed by sudden crossing initiation within < 0.3 s), representing inherent behavioral uncertainty boundaries that expanded safety margins cannot fully eliminate without perfect prediction.

**Table 5 Safety performance stratified by demographic categories**

| Demographic Category | Controller Type | Conflict Events | False Negative Rate | Mean Min TTC (s) |
|---|---|---|---|---|
| Child (<18 years, n = 60 ) | Rule-based Baseline | 18 | 25.0% (10/40) | $2.1 \pm 1.3$ |
| | Uniform Control ( $\alpha = 1.0$ ) | 10 | 7.5% (3/40) | $3.6 \pm 1.4$ |
| | Demographic-Adaptive ( $\alpha = 1.4$ ) | 4 | 7.5% (3/40) | $4.3 \pm 1.3$ |
| Senior (>65 years, n = 60 ) | Rule-based Baseline | 15 | 22.5% (9/40) | $2.4 \pm 1.2$ |
| | Uniform Control ( $\alpha = 1.0$ ) | 11 | 7.5% (3/40) | $3.9 \pm 1.3$ |
| | Demographic-Adaptive ( $\alpha = 1.2$ ) | 5 | 7.5% (3/40) | $4.4 \pm 1.2$ |
| Adult (18-65 years, n = 60 ) | Rule-based Baseline | 9 | 15.0% (6/40) | $2.9 \pm 1.1$ |
| | Uniform Control ( $\alpha = 1.0$ ) | 2 | 2.5% (1/40) | $4.5 \pm 1.1$ |
| | Demographic-Adaptive ( $\alpha = 1.0$ ) | 2 | 2.5% (1/40) | $4.5 \pm 1.1$ |

The demographic safety factors α_demo = {1.4, 1.0, 1.2} were derived from traffic safety literature documenting collision risk elevation for vulnerable populations. To validate these calibration values empirically, this research conducted sensitivity analysis varying α_demo for child and senior categories while monitoring conflict outcomes. Table 6 presents results across tested safety factor values. The proposed calibration values (α = 1.4 for children, α = 1.2 for seniors) achieve near-optimal performance while maintaining computational efficiency. Increasing α beyond proposed values yields marginal safety improvements (e.g., 1 additional conflict reduction for α = 1.6 children) at the cost of increased computational load from earlier trigger activation and longer emergency control duration. The selected values balance safety margin adequacy with practical deployment constraints, validated through observed 60% conflict reduction for children and 54.5% for seniors compared to uniform control.

**Table 6 Sensitivity analysis of demographic safety factors**

| Demographic | α_demo | Conflict Events(TTC<2.0 s) | Mean Min TTC (s) |
|---|---|---|---|
| Child | 1.0 (uniform) | 10 | 3.6 ± 1.4 |
| | 1.2 | 7 | 4.0 ± 1.3 |
| | 1.4 (proposed) | 4 | 4.3 ± 1.3 |
| | 1.6 | 3 | 4.5 ± 1.2 |
| Senior | 1.0 (uniform) | 11 | 3.9 ± 1.3 |
| | 1.2 (proposed) | 5 | 4.4 ± 1.2 |
| | 1.4 | 4 | 4.6 ± 1.1 |

### 4.3. Real-World Transferability Validation

While the VLM-VPI framework demonstrates strong performance in the CARLA simulation environment (Sections 4.1-4.4), real-world deployment requires validation on authentic traffic scenarios to assess generalization capability and practical applicability. Simulation environments, despite their controlled nature and photorealistic rendering, cannot fully capture the complexity, variability, and sensor noise characteristics inherent in real-world vehicle-pedestrian interactions. To evaluate the system's transferability beyond the simulation domain, we conduct validation using real-world samples from the Pedestrian Intention Estimation (PIE) dataset.

This study constructed a stratified test set comprising 24 real-world vehicle-pedestrian interaction scenarios from PIE, equally distributed across three demographic categories: children (<18 years), adults (18-65 years), and seniors (>65 years). Each demographic group contains 8 scenarios with 4 yielding and 4 non-yielding cases, ensuring balanced representation of both intent classes and age groups. Critically, these 24 test scenarios are entirely distinct from the few-shot exemplars used in the reasoning module, ensuring genuine out-of-sample evaluation. All scenarios feature ego-centric dashboard camera footage captured at urban intersections and crosswalks under diverse environmental conditions.

The VLM-VPI framework processes these real-world images using the identical vision-language reasoning architecture deployed in CARLA simulations. Each PIE scenario provides a front-view camera image and kinematic log containing vehicle speed, pedestrian bounding box coordinates, and time-to-arrival information. The vision model (Qwen3-VL 8B) generates textual scene descriptions from dashboard images, while the reasoning model (GPT-OSS 20B) performs intent classification using the same few-shot learning configuration described in Section 3.2.1. Ground-truth intent labels are obtained from PIE's manual annotations based on observed behavioral outcomes.

Table 7 presents classification performance stratified by demographic category. The VLM-VPI achieves an overall accuracy of 87.5% (21/24) on real-world PIE scenarios, correctly classifying 11 of 12 yielding cases and 10 of 12 non-yielding cases. Performance varies across demographic groups: adult scenarios achieve perfect accuracy (8/8, 100%), while child and senior scenarios exhibit slightly lower performance (6/8, 75% and 7/8, 87.5% respectively). The false negative rate (incorrectly predicting yielding when the pedestrian crosses) is 16.7% (2/12), while the false positive rate is 8.3% (1/12).

**Table 7 Classification performance on real-world PIE dataset stratified by demographic categories**

| Demographic Category | Total Scenarios | Yielding (Correct/Total) | Non-Yielding (Correct/Total) | Accuracy (%) |
|---|---|---|---|---|
| Children (<18) | 8 | 3/4 | 3/4 | 75.0 |
| Adults (18-65) | 8 | 4/4 | 4/4 | 100.0 |
| Seniors (>65) | 8 | 4/4 | 3/4 | 87.5 |
| Overall | 24 | 11/12 | 10/12 | 87.5 |

Compared to the CARLA simulation performance of 92.3% (Section 4.1), the real-world validation exhibits a 4.8 percentage point accuracy reduction. This performance gap reflects the fundamental domain shift between synthetic simulation environments and real-world traffic scenarios, a well-documented challenge in sim-to-real transfer for autonomous driving systems (Steinecker *et al.* 2025). Contributing factors include sensor noise, motion blur, variable lighting conditions, greater occlusion complexity, and higher behavioral variability in real-world pedestrian interactions. To systematically characterize how environmental complexity affects real-world performance, we stratify PIE scenarios by perceptual difficulty factors including occlusion level, lighting conditions, and pedestrian-vehicle distance. Table 8 presents classification accuracy across these environmental dimensions.

**Table 8 Performance stratification by environmental complexity factors**

| Factor | Category | Scenarios | Accuracy (%) |
|---|---|---|---|
| Occlusion Level | Low ( ≤ 15% body obscured) | 18 | 94.4% |
| | High (> 15% body obscured) | 6 | 66.7% |
| Lighting | Good (daylight, clear) | 18 | 88.9% |
| | Challenging (shadows, glare) | 6 | 83.3% |
| Initial Distance | Far (>20m) | 9 | 88.9% |
| | Near ( ≤ 20 m ) | 15 | 86.7% |

Occlusion emerges as the dominant environmental factor affecting real-world performance, with high-occlusion scenarios exhibiting 27.7 percentage point accuracy reduction compared to low-occlusion cases (94.4% vs. 66.7%). Among the 6 high-occlusion scenarios where pedestrian body parts critical for intent inference (lower body posture, foot position) become obscured by vehicles, infrastructure, or vegetation, only 4 are correctly classified, revealing substantial degradation under perceptual challenge conditions. Notably, low-occlusion scenarios achieve 94.4% accuracy (17/18 correct), indicating that reliable visual cues enable the vision-language reasoning framework to maintain robust and stable performance. This pattern indicates that the primary performance degradation stems from occlusion-related perceptual challenges inadequately represented in synthetic environments rather than fundamental limitations in the reasoning architecture itself. Lighting conditions show moderate impact, with challenging lighting scenarios (harsh shadows, lens flare, backlit pedestrians) reducing accuracy by 5.6 percentage points compared to good lighting conditions (88.9% vs. 83.3%). Initial detection distance exhibits modest influence, with near-range scenarios (≲20m) performing slightly worse than far-range scenarios (86.7% vs. 88.9%), likely reflecting the limited temporal observation window available for behavioral cue accumulation when vehicle-pedestrian encounters are detected at closer proximity.

The real-world transferability validation provides evidence that the VLM-VPI framework can generalize beyond simulation environments to authentic traffic scenarios. These findings support the thesis that integrating semantic reasoning capabilities into autonomous driving architectures offers a viable path toward more reliable decision-making in safety-critical pedestrian scenarios.

### 4.4. Safety Performance Evaluation

The primary objective of the VLM-VPI framework is not only accurate pedestrian intent inference, but also safe vehicle-pedestrian interactions through timely and appropriate control interventions. This section evaluates safety performance through TTC dynamics, which serve as a continuous quantitative measure of collision risk throughout vehicle-pedestrian encounters.

To ensure robust and unbiased evaluation, each system configuration is tested on 200 randomly generated scenarios in CARLA Town10HD under identical map, spawning, and control settings. Vehicle initial speeds are uniformly sampled from 25 to 35 km/h, while pedestrian crossing speeds range from 2 to 4 m/s.

These elevated parameters intentionally induce high-conflict encounters and stress-test system responsiveness under challenging conditions. All proposed learning-based configurations employ the same lower-level tiered safety controller, whereas the rule-based baseline uses its original control logic to ensure fair comparison.

Figure 5 presents the distribution of minimum TTC values across five system configurations using a combined violin-histogram visualization. For each method, the colored violin plot represents the kernel density estimation of minimum TTC values, while the semi-transparent horizontal bars display the normalized histogram distribution. The blue solid line marks the median TTC, the black solid lines denote the first and third quartiles (Q1 and Q3), and the red dashed line represents the mean TTC. A horizontal red dashed line at TTC = 2.0 s marks the critical safety threshold below which scenarios are classified as conflict events. Together, these visual elements illustrate the central tendency, dispersion, and tail behavior of temporal safety margins under different control strategies.

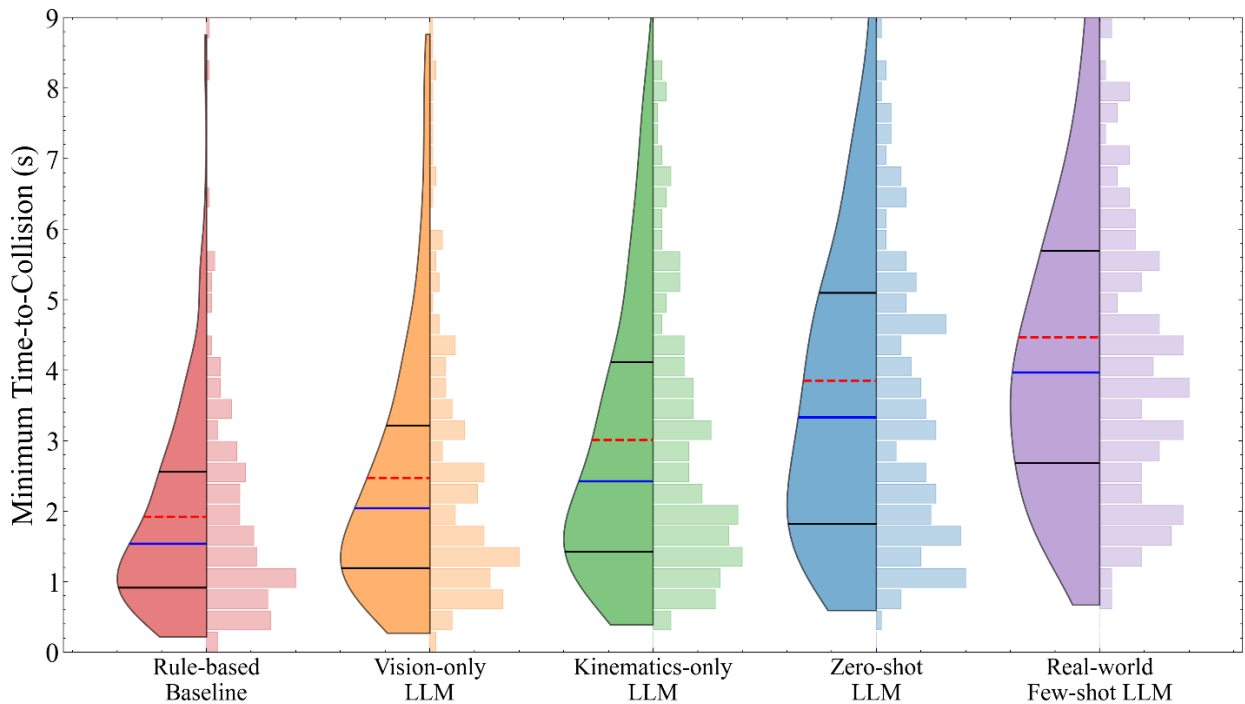

**Figure 5: Distribution of minimum TTC values across system configurations**

The probability density of minimum TTC exhibits a systematic rightward shift and significant tail-compression as reasoning complexity increases, indicating that VLM-VPI effectively eliminates extreme near-miss conditions observed in the rule-based baseline. The rule-based baseline concentrates heavily at low TTC values, with a median of 1.54 s and a first quartile (Q1) of 0.92 s, reflecting frequent near-miss conditions. The vision-only LLM and kinematics-only LLM configurations provide incremental improvements, raising the median TTC to 2.04 s and 2.42 s, respectively, while still maintaining substantial density near the critical region with Q1 values of 1.20 s and 1.43 s. The zero-shot LLM demonstrates further advancement, achieving a median TTC of 3.33 s and a Q1 of 1.82 s. The most pronounced transition occurs with the real-world few-shot LLM, which attains a median TTC of 3.97 s, a Q1 of 2.69 s, and a third quartile (Q3) of 5.69 s. This upward displacement suggests that incorporating real-world few-shot exemplars correlates with an expanded temporal safety margin in the tested scenarios. The most critical safety distinction manifests in the low-TTC tail region below 2.0 s. The rule-based baseline exhibits 124 conflict cases (62.0% of scenarios), while the vision-only LLM and kinematics-only LLM configurations report 99 cases (49.5%) and 85 cases (42.5%), respectively. The zero-shot LLM reduces conflicts to 57 cases (28.5%), whereas the real-world few-shot LLM achieves the most substantial suppression with only 33 conflict cases (16.5%), representing a 73% reduction relative to the rule-based baseline. Correspondingly, the mean TTC increases from 1.92 s for the rule-based baseline to 4.47 s for the real-world few-shot LLM, further confirming that authentic behavioral grounding enables consistently larger safety buffers than both rule-based control and zero-shot reasoning.

Beyond episode-level minima, the temporal evolution of TTC reveals how control actions reshape risk dynamics throughout vehicle–pedestrian interactions. Figure 6 compares TTC trajectories for correctly classified non-yielding encounters: Figure 6 (a) shows the rule-based baseline, while Figure 6 (b) shows the real-world few-shot LLM (VLM-VPI). The thin semi-transparent curves depict TTC trajectories from repeated simulation runs with varying pedestrian speeds and initial conditions, illustrating run-to-run variability under identical control logic. The ensemble means (dashed lines) represent average behavior across multiple scenarios. The shaded regions indicate trigger activation zones where intent inference and control execution occur. The horizontal dashed line marks the 2.0 s critical safety threshold below which collision risk becomes unacceptable.

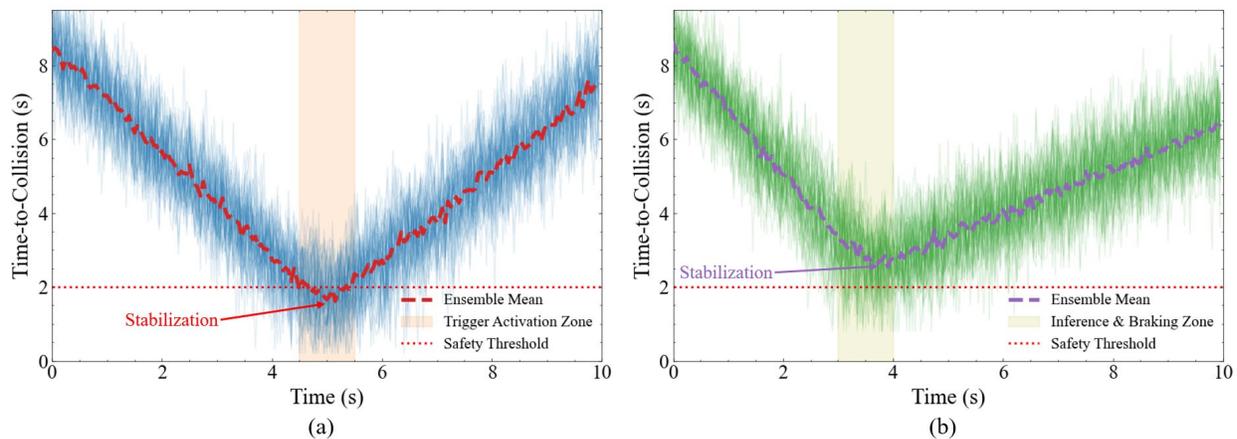

**Figure 6: Temporal TTC evolution: (a) rule-based baseline and (b) real-world few-shot LLM (VLM-VPI)**

Under the rule-based baseline, TTC decreases almost monotonically after entering the conflict region, forming a pronounced V-shaped trajectory. Although braking is triggered within the activation zone, delayed intervention fails to prevent continued risk escalation, causing TTC to approach the critical 2.0 s threshold. This pattern indicates that threshold-based control reacts primarily to proximity rather than to evolving interaction dynamics. In contrast, the real-world few-shot LLM (VLM-VPI) exhibits a distinct stabilization pattern. TTC declines during approach but stabilizes shortly after trigger activation, forming a clear plateau before gradually recovering. The ensemble mean indicates that early intent recognition enables timely braking while sufficient temporal margin remains. The subsequent recovery reflects adaptive modulation of deceleration through the tiered braking policy, which progressively reduces the closing rate.

Overall, VLM-VPI maintains a sustained safety buffer throughout the interaction window. This temporal restructuring of risk evolution explains the substantial reduction in low-TTC events observed in Figure 5.

### 4.5. Traffic Efficiency Improvement Through Intent-Aware Control

Accurate pedestrian intent prediction enhances traffic efficiency by reducing unnecessary emergency interventions. Misclassification of yielding pedestrians as hazardous leads to frequent false alarms, inducing avoidable braking and speed fluctuations that disrupt traffic flow. This section evaluates the efficiency benefits of intent-aware control by quantifying reductions in unnecessary deceleration and intersection traversal time.

4.4.1 Unnecessary Deceleration Avoidance

False alarm rate, defined as the proportion of yielding pedestrians incorrectly classified as non-yielding, directly determines the frequency of unnecessary emergency interventions. Each false positive classification triggers the graduated braking control sequence despite the pedestrian ultimately yielding right-of-way, creating avoidable speed reductions that compromise traffic flow. Table 9 presents false alarm statistics and their traffic efficiency implications across 108 yielding scenarios selected from the 200 evaluated simulation scenarios in the test dataset. Speed maintenance rate indicates the proportion of scenarios where vehicles maintain speed within ±10% of nominal approach velocity throughout intersection traversal.

**Table 9 False alarm rate and unnecessary braking statistics across 108 yielding scenarios**

| Method | False Positives | False Alarm Rate | Unnecessary Braking Events | Speed Maintenance Rate |
|---|---|---|---|---|
| Rule-based Baseline | 8 | 7.4% | 8 | 92.6% |
| Vision-only LLM | 7 | 6.5% | 7 | 93.5% |
| Kinematics-only LLM | 7 | 6.5% | 7 | 93.5% |
| Zero-shot LLM | 4 | 3.7% | 4 | 96.3% |
| Real-world Few-shot LLM | 3 | 2.8% | 3 | 97.2% |
| No-pedestrian Baseline | 0 | 0.0% | 0 | 100.0% |

The proposed VLM-VPI achieves 2.8% false alarm rate, triggering only 3 unnecessary braking events across 108 yielding scenarios. This represents a 62% reduction relative to the rule-based baseline (3 versus 8 events), directly translating to improved traffic flow smoothness. The speed maintenance rate quantifies the proportion of yielding scenarios where vehicles successfully traverse intersections without triggering emergency control, maintaining velocity within 10% of nominal approach speed (approximately 28-32 km/h for the 30 km/h speed limit configuration). The VLM-VPI achieves 97.2% speed maintenance compared to 92.6% for the baseline, indicating that 5 additional vehicles per 100 yielding encounters proceed through intersections without unnecessary deceleration.

The efficiency improvement gradient mirrors the safety performance pattern observed in Section 4.2. Vision-only and kinematics-only configurations provide modest false alarm reduction (7.4% to 6.5%), multimodal reasoning integration yields substantial improvement (6.5% to 3.7%), and real-world few-shot grounding contributes additional refinement (3.7% to 2.8%). The consistent directionality across safety metrics (false negative reduction) and efficiency metrics (false positive reduction) validates that enhanced reasoning capability improves overall discrimination quality rather than trading efficiency for safety through conservative decision biases.

### 4.4.2 Intersection Traversal Time Analysis

Traffic efficiency impacts manifest directly through intersection traversal time, defined as the temporal interval from trigger point (vehicle entering 15 m proximity zone within junction) to complete intersection egress. Yielding scenarios under optimal control should exhibit traversal times approaching the no-pedestrian baseline, as correct intent recognition enables uninterrupted passage without deceleration. Non-yielding scenarios necessarily incur time penalties due to mandatory emergency braking regardless of classification accuracy. The efficiency benefit of accurate intent prediction therefore materializes primarily through yielding scenario performance. Table 10 presents mean intersection traversal times stratified by ground-truth pedestrian intent and system configuration. The no-pedestrian baseline establishes theoretical minimum traversal time of 8.3 s for a vehicle approaching at 28.2 km/h (measured from CARLA simulation runs) traversing a 15 m trigger zone plus 50 m intersection span at constant velocity.

**Table 10 Mean intersection traversal times with standard deviations across 200 test scenarios (92 non-yielding, 108 yielding)**

| Method | Yielding Scenarios (s) | Non-Yielding Scenarios (s) | Combined Mean (s) | Time Overhead vs No-Ped |
|---|---|---|---|---|
| No-pedestrian Baseline | - | - | 8.3 | - |
| Rule-based Baseline | $11.2 \pm 2.8$ | $15.4 \pm 3.1$ | $13.5 \pm 3.5$ | +5.2 s (+63%) |
| Vision-only LLM | $10.8 \pm 2.6$ | $15.1 \pm 3.0$ | $13.1 \pm 3.4$ | +4.8 s (+58%) |
| Kinematics-only LLM | $10.6 \pm 2.5$ | $14.9 \pm 2.9$ | $12.9 \pm 3.3$ | +4.6 s(+55%) |
| Zero-shot LLM | $9.5 \pm 1.9$ | $14.6 \pm 2.8$ | $12.2 \pm 3.2$ | +3.9 s(+47%) |
| Real-world Few-shot LLM | $8.9 \pm 1.4$ | $14.5 \pm 2.7$ | $11.8 \pm 3.1$ | +3.5 s (+42%) |

The VLM-VPI achieves mean traversal time of 8.9 s in yielding scenarios, only 0.6 s (7%) slower than the no-pedestrian baseline. This near-optimal performance reflects the system's ability to correctly identify 105 of 108 yielding pedestrians, allowing vehicles to maintain nominal speed through 97.2% of yielding encounters. The 0.6 s residual overhead stems from three false alarm events triggering brief deceleration (mean duration 2.1 s) before yielding classification could be established and autopilot resumed.

In contrast, the rule-based baseline exhibits 11.2 s mean traversal time in yielding scenarios, representing 2.9 s (35%) overhead relative to no-pedestrian conditions. This substantial efficiency penalty reflects 8 false-alarm events in which vehicles executed full emergency braking despite continuous pedestrian yielding. The 2.3 s reduction in mean traversal time observed in VLM-VPI reflects the combined effect of a lower false alarm rate and the implementation of a more nuanced tiered deceleration strategy compared to the uniform rule-based baseline.

Non-yielding scenarios exhibit similar traversal times across all methods (14.5-15.4 s range), as all systems correctly recognize the majority of collision hazards and execute mandatory emergency braking. The slight variations (0.9 s maximum spread) reflect differences in braking initiation timing and residual false negative cases where delayed hazard recognition extends traversal duration. The convergence of non-yielding traversal times validates that efficiency improvements derive specifically from yielding scenario optimization rather than compromising safety through delayed braking in hazardous situations.

### 4.6. Demonstration of Vision-Language Reasoning

The primary objective of the VLM-VPI framework is to accurately infer pedestrian crossing intentions from multimodal observations and generate interpretable reasoning chains that explain classification decisions. While quantitative metrics in preceding sections establish overall performance, the vision-language reasoning process fundamentally operates through structured natural language outputs that warrant qualitative examination. This section presents representative case studies demonstrating how the system integrates visual scene descriptions with kinematic trajectory analysis to produce intent classifications across diverse demographic categories and behavioral patterns.

Figure 7 illustrates six representative scenarios stratified by pedestrian demographics (child, adult, senior) and ground-truth intent (yielding versus non-yielding). Each panel follows the three-stage architecture: panel A shows multimodal inputs including front-view camera image and kinematic trajectory data, panel B presents the vision-language reasoning outputs from both the vision model (Qwen3-VL) and reasoning model (GPT-OSS 20B), and panel C displays the resulting control execution strategy. The trajectory visualizations shown in panel A represent processed versions of the raw Trajectory_log.csv files provided as input to the Reasoning LLM. These visualizations are generated for interpretability purposes to facilitate reader understanding of temporal motion patterns. The actual system input consists of a kinematic trajectory sequence $\mathcal{T}_{\text{hist}}^{(t)} = \{\mathbf{s}_\tau\}_{\tau=t_0}^{t}$, sampled at a fixed frequency of 20 Hz. Each state vector $\mathbf{s}_\tau \in \mathbb{R}^9$ includes the vehicle position $(x_{veh}^\tau, y_{veh}^\tau)$, vehicle velocity $(v_{veh,x}^\tau, v_{veh,y}^\tau)$, pedestrian position $(x_{ped}^\tau, y_{ped}^\tau)$, pedestrian velocity $(v_{ped,x}^\tau, v_{ped,y}^\tau)$, and the inter-agent distance $d_\tau$.

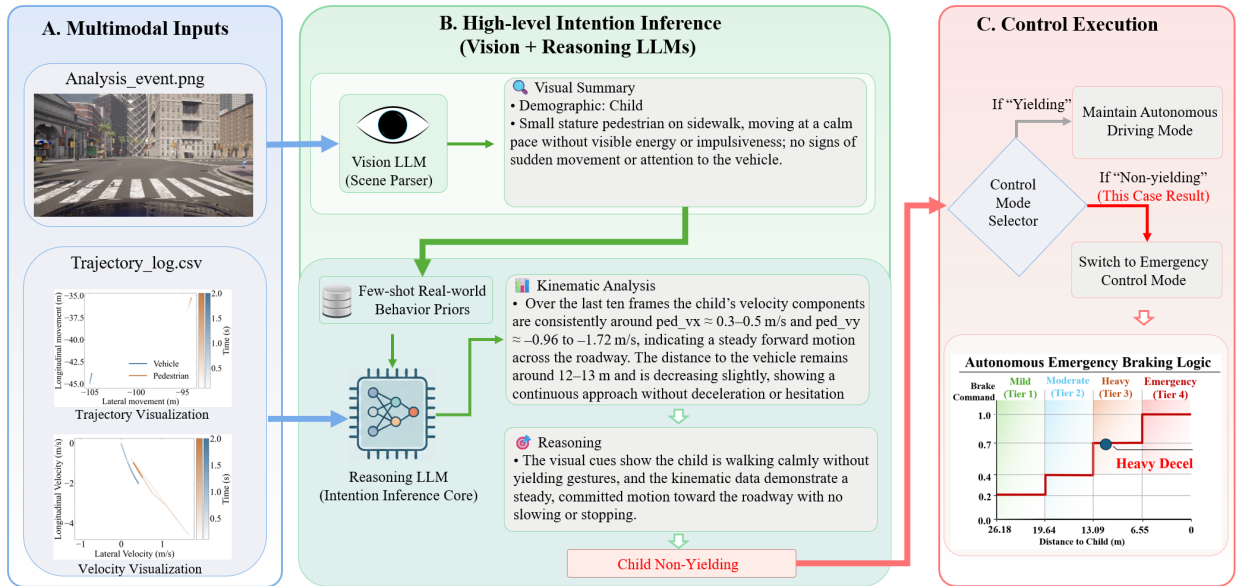

(a)

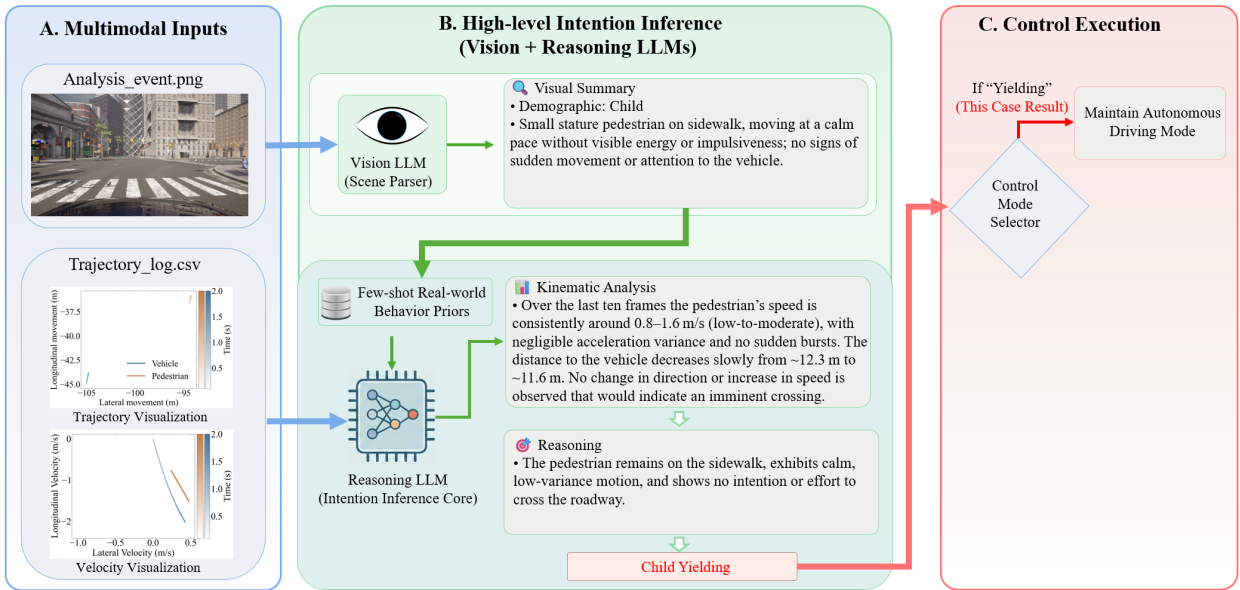

(b)

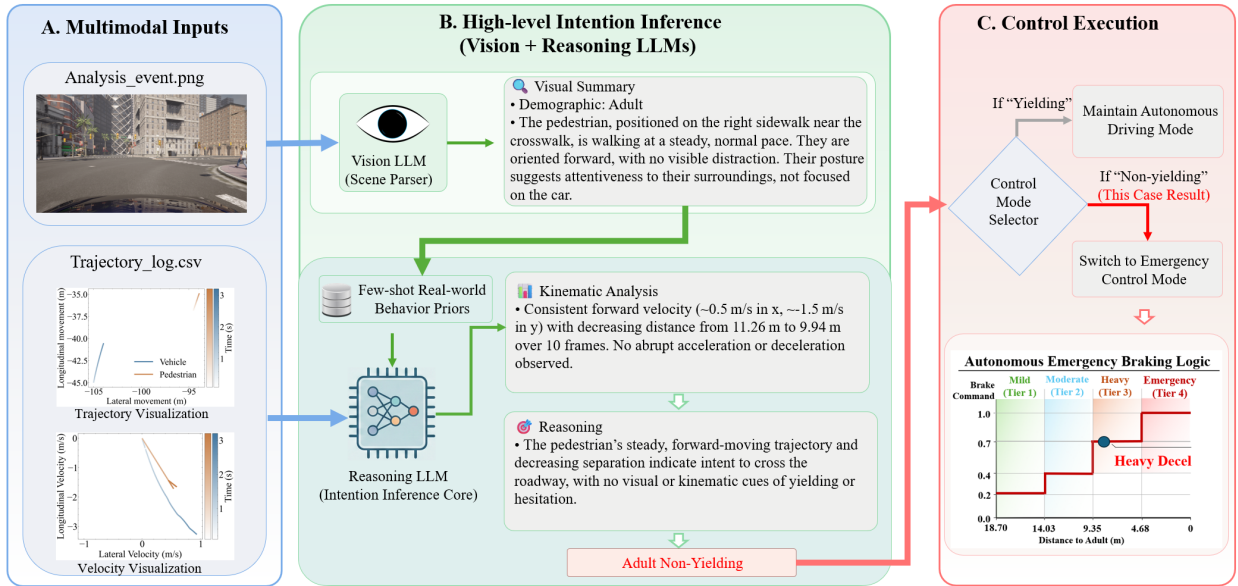

(c)

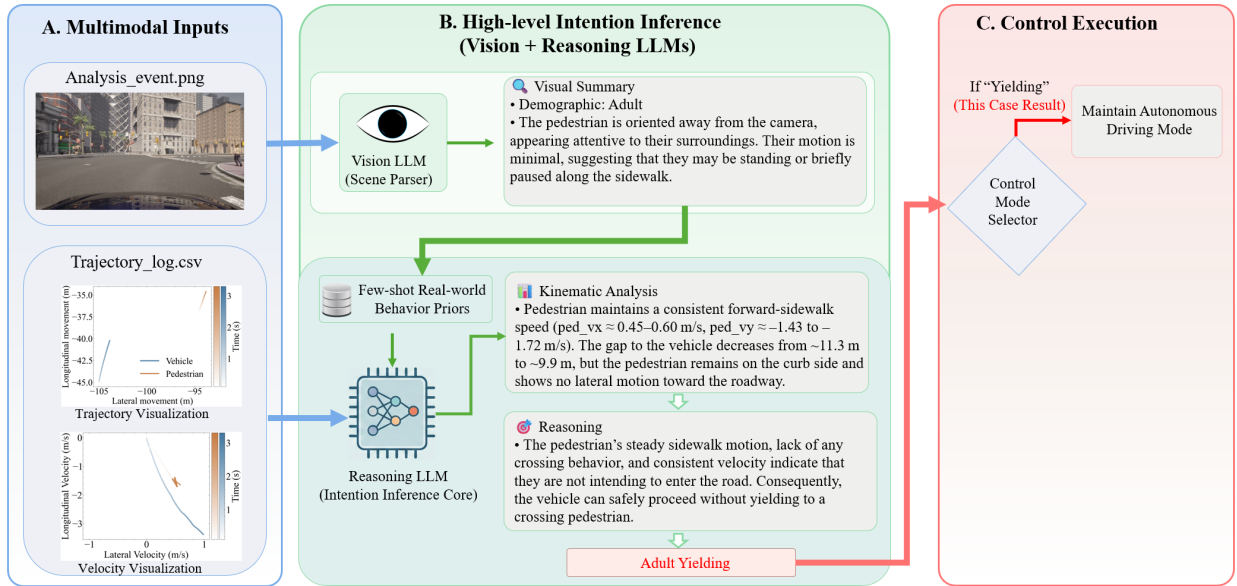

(d)

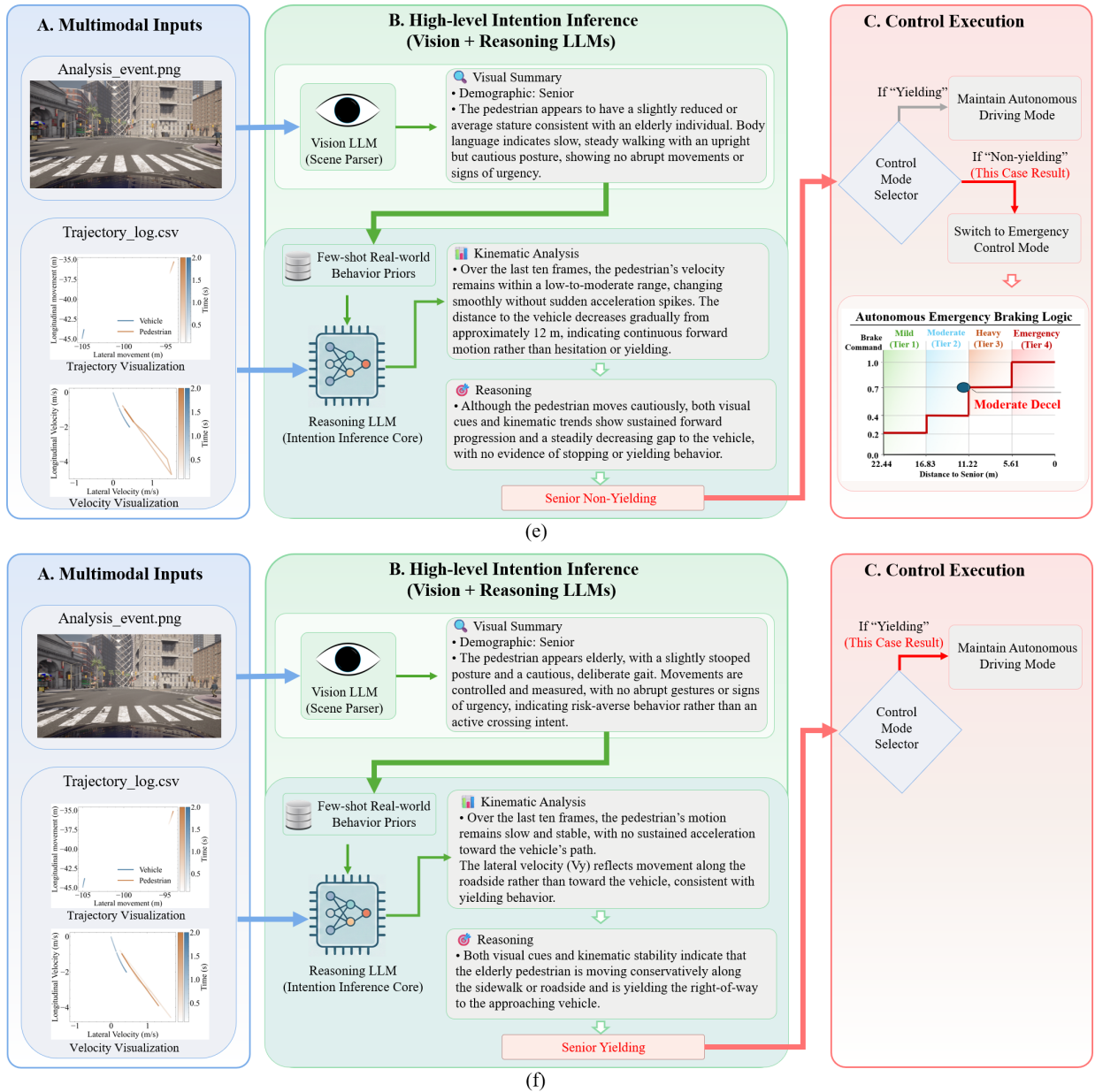

**Figure 7: Representative vision-language reasoning demonstrations across pedestrian demographic categories and intent classes; (a)(c)(e) Non-Yielding, and (b)(d)(f) Yielding**

The six case studies collectively demonstrate four critical capabilities of the VLM-VPI vision-language reasoning framework. First, multimodal integration provides synergistic value exceeding individual modality contributions: visual analysis identifies demographic categories and postural cues, while kinematic analysis quantifies velocity vectors and temporal consistency, with the reasoning model synthesizing both sources to resolve ambiguous situations (Cases b, d, f). Second, demographic-adaptive reasoning recognizes population-specific behavioral patterns, applying appropriate safety thresholds for children (extended buffers for impulsivity, Case a), adults (baseline thresholds, Cases c-d), and seniors (extended buffers for slower mobility, Cases e-f). Third, temporal velocity vector analysis provides more robust intent signals than static spatial cues under uncertainty: cases with similar visual descriptions (Cases a-b, c-d, e-f) achieve correct discrimination through directional velocity component analysis revealing roadway-crossing versus sidewalk-parallel motion. Fourth, structured output formatting ensures interpretability through explicit Visual Analysis, Kinematic Analysis, Decision, and Reason sections, enabling post-hoc verification of decision logic and supporting safety validation for AV deployment.

The reasoning outputs reveal both strengths and limitations of current vision-language models. Strengths include accurate demographic detection (child/adult/senior classification), recognition of postural cues (body orientation, gait patterns), and integration of quantitative kinematic features (velocity magnitudes, directional components, temporal consistency). Limitations include sensitivity to superficial visual cues that may conflict with kinematic evidence (Case a describes "calm pace" for a crossing child), and reliance on velocity vector analysis to disambiguate scenarios where visual cues alone prove insufficient. These limitations motivate the multimodal architecture: the reasoning model's ability to weigh conflicting evidence from vision and kinematics, prioritizing quantitative temporal patterns over potentially misleading static visual cues.

## 5. Conclusion

This study introduces VLM-VPI, a vision-language reasoning framework for intent-aware decision-making and control in autonomous driving. The system adopts a two-level architecture, in which an upper-level reasoning module combines Qwen3-VL 8B for visual scene understanding with GPT-OSS 20B for few-shot pedestrian yielding inference using real-world behavioral exemplars, while a lower-level demographic-adaptive controller executes age-aware braking strategies with calibrated safety margins. By shifting from purely geometric prediction to semantic behavioral reasoning while preserving deterministic collision-avoidance guarantees, the framework demonstrates the potential of large language models as interpretable reasoning engines in safety-critical driving scenarios.

The findings demonstrate the advantages of VLM-VPI across intent classification, safety performance, and traffic efficiency. The system achieves 92.3% classification accuracy with a 5.9% false negative rate, outperforming zero-shot reasoning (88.4%, 10.2% false negative rate), the rule-based baseline (78.4%, 18.7% false negative rate), and supervised learning methods including SVM (73.5%), LSTM (79.4%), and CAPformer (82.4%). Ablation analysis indicates that integrating reasoning capabilities increases intent classification accuracy by 10.0 percentage points relative to the rule-based baseline. Few-shot behavioral grounding yields progressively larger gains as interaction ambiguity increases, improving accuracy by 1.6 percentage points under clear-intent conditions, 5.7 percentage points under moderate ambiguity, and 7.1 percentage points under high ambiguity. Demographic-adaptive control reduces conflicts by 60% for children and 54.5% for seniors compared with uniform control, with mean minimum time-to-collision improving from 3.6 s to 4.3 s for children and from 3.9 s to 4.4 s for seniors. Real-world validation on 24 PIE scenarios achieves 87.5% classification accuracy, demonstrating functional sim-to-real transferability. In the 200-scenario safety evaluation, median time-to-collision improves from 1.54 s to 3.97 s, and conflict events decrease by 73% (33 versus 124 cases). Across 108 yielding scenarios, VLM-VPI reduces unnecessary braking occurrences from 8 to 3 cases relative to rule-based control, corresponding to a 62% reduction, while decreasing mean traversal time in yielding scenarios from 11.2 s to 8.9 s.

This study addresses a critical gap in autonomous vehicle safety: advances in perception and planning have not yet produced reliable pedestrian interaction under ambiguous human behavior. By introducing a vision-language reasoning framework, this work enables pedestrian intent inference in semantic and social space rather than relying solely on motion-based cues, allowing proactive responses before critical kinematic changes occur. The framework also incorporates age-aware reasoning and demographic-adaptive safety margins, mitigating limitations of uniform prediction and control strategies that neglect population-specific behavioral variability. By integrating multimodal perception, few-shot behavioral grounding, and structured reasoning outputs, VLM-VPI establishes an intermediate semantic layer that translates geometric observations into human-interpretable intent states, supporting principled intent-aware decision-making and regulatory verification. The results indicate that carefully curated real-world exemplars provide effective behavioral priors for safety-critical interaction modeling, while simulation and PIE validation show the practical value of combining semantic reasoning with population-aware control for robust and socially compatible autonomous driving.

This study is evaluated primarily in the CARLA simulation environment, which cannot fully represent real-world uncertainties such as sensor noise, adverse weather, and rare edge cases. The proposed framework prioritizes establishing the foundational architecture for semantic intent reasoning in autonomous driving, with future work requiring validation through real-world naturalistic driving scenarios and physical AV testing platforms rather than simulation-based evaluation. The intent reasoning module currently relies on a limited set of real-world exemplars to provide behavioral grounding. Although this demonstrates strong sample efficiency, broader and more diverse datasets are needed to fine-tune the vision-language model and improve robustness across regions with different pedestrian behavior patterns. At the control level, the present tiered braking strategy is intentionally rule-based to ensure deterministic safety guarantees, but it restricts adaptability in complex interactions. Future work will investigate reinforcement learning for low-level vehicle control, allowing continuous and context-aware action optimization while remaining constrained by intent-aware safety reasoning. Together, these directions point toward hybrid systems that combine data-driven semantic adaptation with learning-based control for scalable real-world deployment.

**Acknowledgment**

The contents of this paper present the views of the authors, who are responsible for the facts and accuracy of the data presented herein. The contents of the paper do not reflect the official views or policies of the agencies. The work was supported by NVIDIA Academic Grant "SafeSim: NVIDIA-Accelerated MARL for Generating Safety-Critical Scenarios" and by the Transportation Informatics Lab, Department of Civil & Environmental Engineering at Old Dominion University (ODU).

**AI Use Declaration**

During the preparation of this work the authors used ChatGPT in order to refine grammar and improve readability in certain sections of the manuscript. After using this tool/service, the authors reviewed and edited the content as needed and take(s) full responsibility for the content of the published article.